\documentclass[a4paper]{jpconf}
\usepackage{graphicx}
\usepackage{citesort}
\usepackage{amsmath,amssymb}
\begin{document}
\title{Statistical properties of an ideal subgrid-scale correction for Lagrangian particle tracking in turbulent channel flow}

\author{F. Bianco$^{1,2}$, S. Chibbaro$^2$, C. Marchioli$^1$, M.V. Salvetti$^3$, A. Soldati$^1$}

\address{$^1$Energy Technology Dept.and
Centro Interdipartimentale di Fluidodinamica e Idraulica,
University of Udine, 33100 Udine -- Italy}
\address{$^2$ Jean Le Rond D'Alembert Institute, 
University Pierre et Marie Curie and CNRS UMR7190,  
75252 Paris Cedex 05 --  France}
\address{$^3$Aerospace Engineering Dept., University of Pisa, 56100 Pisa -- Italy}

\ead{mv.salvetti@dia.unipi.it}

\begin{abstract}
One issue associated with the use of Large-Eddy Simulation (LES) to investigate the dispersion of small inertial particles in turbulent flows is the accuracy with which particle statistics and concentration can be reproduced. The motion of particles in LES fields may differ significantly from that observed in experiments or direct numerical simulation (DNS) because the force acting on the particles is not accurately estimated, due to the availability of the only filtered fluid velocity, and because errors accumulate in time leading to a progressive divergence of the trajectories. This may lead to different degrees of inaccuracy in the prediction of statistics and concentration. We identify herein an ideal subgrid correction of the a-priori LES fluid velocity seen by the particles in turbulent channel flow. This correction is computed by imposing that the trajectories of individual particles moving in filtered DNS fields exactly coincide with the particle trajectories in a DNS.
In this way the errors introduced by filtering into the particle motion equations can be singled out and analyzed separately from those due to the progressive divergence of the trajectories. The subgrid correction term, and therefore the filtering error, is characterized in the present paper in terms of statistical moments. The effects of the particle inertia and of the filter type and width on the properties of the correction term are investigated.
\end{abstract}

\section{Introduction}

Dispersion and entrainment of inertial heavy particles in turbulent flows are crucial in a number of industrial applications and environmental phenomena. Examples are mixing, combustion, depulveration, spray dynamics, pollutant dispersion or cloud dynamics. The dynamics of inertial particles in turbulent flows is characterized by large-scale clustering and preferential concentration, due to the tendency of inertial particles to distribute preferentially at the periphery of strong vortical regions and to segregate into straining regions \cite{hit2,ms02,hit1,fs06,re01}. In turbulent boundary layers clustering and preferential concentration also influence settling, deposition and entrainment to and from the wall \cite{ms02,re01,Soldati2009}. 

Direct numerical simulation (DNS) together with Lagrangian particle tracking (LPT) have successfully been used to investigate and quantify the behavior of particles in wall-bounded turbulent flow. DNS studies have shown, for instance, that particle segregation and subsequent deposition phenomena are strictly related with the interaction between particles and the near-wall coherent vortical structures \cite{ms02,pms05,Soldati2009}. More specifically, pairs of counter-rotating quasi-streamwise vortices, which are the statistically most common near-wall structures, are able to entrain particles from the core region towards the wall on their downwash side and away from the wall on their upwash side. This transport mechanism, called turbophoresis \cite{r83}, produces net particle fluxes toward the wall leading to irreversible particle accumulation at the wall.

On the other hand, DNS is still nowadays limited to low Reynolds numbers, often very far from those characterizing practical applications, and to  academic flow configurations. Conversely, thanks to the huge growth of available computational resources, large-eddy simulation (LES) can tackle increasingly complex problems. In LES the basic idea is to directly simulate only the turbulence scales larger than a given dimension, while the effects of the unresolved subgrid scales (SGS) on the large-scale motion are modeled. From a formal viewpoint, the separation between scales is provided through filtering, which can be explicitly applied to the flow equations or implicitly introduced by the numerical discretization. Thus, the exact dynamics of the filtered fluid velocity field can at best be obtained from LES (\textit{ideal LES}), this when SGS modeling and numerical errors are negligible. Let us restrict ourselves to the context of an ideal LES, which can be mimicked in the so-called a-priori tests, i.e. by assuming that the LES velocity fields are represented by the filtered DNS ones. In LPT coupled with LES of the fluid phase, the forces acting on particles can be estimated from the filtered fluid velocity fields, by simply neglecting the effects of the SGS scales. A critical issue is, however, to evaluate the impact of this approximation on the ability to capture the previously described phenomena characterizing the particle dynamics, i.e. preferential concentration and turbophoresis, and to accurately predict quantities of practical interest, such as particle concentration profiles. 

In the last decades, several examples of LPT coupled with LES of wall-bounded turbulent flows can be found in the literature \cite {uo96,ws96,armenio,kv05,k06,Yamamoto2001,Marchioli2006,Marchioli2008,Vreman2009,Sengupta2009,Khan2010}. In most of the earlier studies, the fluid SGS velocity fluctuations were simply neglected in the particle motion equations \cite{uo96,ws96,armenio,Yamamoto2001,Marchioli2006}. This is based on the assumption that, if the particle response time is large enough compared to the smallest time scale resolved in the LES, the effects of the SGS fluid velocity fluctuations on the particle velocity statistics are negligible \cite{armenio,fs06}. For well-resolved LES, this assumption holds to capture
satisfactorily the statistics of particle velocity. Nonetheless, more recent studies \cite{kv05,k06,Marchioli2008,Khan2010} have demonstrated that neglecting the effects of the fluid SGS velocity fluctuations on particle motion leads to significant underestimation of preferential concentration and, consequently, to weaker deposition fluxes and lower near-wall accumulation. The issue is therefore how to model the SGS effects in the particle motion equations.  

A few different ways have been explored in the literature. Filter inversion or approximate deconvolution have been used in LPT in homogeneous isotropic turbulence \cite{sm05,Gobert2010} and in turbulent channel flow \cite{kv05,k06,Marchioli2008b}. These techniques appeared to be successful in recovering the correct fluid energy content for the resolved scales and improved the prediction of preferential concentration and turbophoresis\cite{kv05,k06,Marchioli2008b}. Nonetheless, these models still cannot recover the effects of the unresolved SGS scales and previous studies \cite{Marchioli2008b} show that the reintroduction of the correct amount of fluid velocity fluctuations does not warrant accurate quantitative prediction of particle preferential concentration and near-wall accumulation and suggest that additional information on the SGS flow field is required. Attempts of introducing information on the unrepresented scales have been made through fractal interpolation \cite{Marchioli2008b} or kinematic simulation\cite{Khan2010}. Stochastic models have also been proposed for LPT coupled with LES of the carrier phase, but, they have mainly been used for homogeneous isotropic turbulence \cite{sm06,Pozorski2009,Gobert2010}. 

In order to devise efficient SGS models for LPT, the quantification and the analysis of the errors introduced by the lack of SGS scales on particle dynamics is clearly a necessary step. Even if we restrict ourselves to ideal LES, i.e. we assume that the LES of the fluid phase provides the exact dynamics of the resolved scales, this task is complicated by the fact that there are still two sources of errors. The first one, is the inaccuracy in the estimation of the forces acting on particles when they are computed only from the filtered fluid velocity, output of the LES of the carrier phase. This can be considered as the pure filtering error. The second contribution to the error comes from the fact that, due to the inaccurate estimation of forces, considering the same initial distribution of particles, the trajectories obtained in LES progressively diverge from the \textit{exact} DNS ones and thus the forces are evaluated at different locations. When comparing particle concentrations and statistics obtained in LES to the DNS reference data, the effects of the two errors can not easily be singled out. In a recent work \cite{Calzavarini2010} the two contributions have been computed separately for couples of tracer particles in a-priori LES of homogeneous isotropic turbulence. 

The main goal of the present paper is to compute a SGS correction term for the LES fluid velocity seen by inertial particles, which accounts for the pure filtering error alone. This correction is computed by assuming that the trajectories of particles having the same initial position tracked in a-priori LES coincide with the ones obtained in DNS of turbulent channel flow. This is clearly an idealized situation and is a stronger requirement with respect to what is usually asked to LPT coupled to LES, i.e. to correctly reproduce the particle statistics and concentration. Nevertheless, in this way, the effects of the pure filtering error can be singled out and quantified, since the divergence of trajectories is a-priori prevented. This \textit{ideal} correction (IC) can be computed for each single particle and at each time step along its trajectory, and the properties of IC can be characterized in terms of statistics and probability distribution function. In the present paper, the correction term is characterized in terms of statistical moments, while PDFs will be presented in forthcoming studies. The aim is to obtain indications of the key features to be incorporated in SGS models for the particle motion equations. The IC term is computed here for different filter widths and types, corresponding roughly to varying amounts of resolved flow energy, and for different particle inertia. A point at issue is indeed whether SGS models for LPT should explicitly take into account inertia effects or general conclusions may be drawn. Finally, an additional question is whether the flow inhomogeneity has an impact on the key features of IC and therefore whether a SGS model should have different characteristics in the near-wall region than in the center of the channel. 

\section{Physical Problem and Methodology} 
\label{meth}

\subsection{Particle-laden turbulent channel flow and DNS methodology}

The flow into which particles are introduced is a turbulent
channel flow of gas. The reference geometry consists of two infinite
flat parallel walls: the origin of the coordinate system is located
at the center of the channel and the $x-$, $y-$ and $z-$ axes point
in the streamwise, spanwise and wall-normal directions respectively.
Periodic boundary conditions are imposed on the fluid velocity field
in $x$ and $y$, no-slip boundary conditions
are imposed at the walls. In the present study, we consider air (assumed
to be incompressible and Newtonian)
with density $\rho = 1.3~kg~m^{-3}$ and kinematic
viscosity $\nu = 15.7{\times}10^{-6}~m^{2}~s^{-1}$.
The flow is driven by a mean dimensionless pressure gradient, 
such that the shear Reynolds number, $Re_{\tau} = u_{\tau}h/\nu$, based on the shear (or friction) velocity, $u_{\tau}$,
and on the half channel height, $h$, is equal to 150.
The shear velocity is defined as $u_{\tau} = (\tau_{w}/\rho)^{1/2}$,
where $\tau_{w}$ is the mean shear stress at the wall.

Particles with density $\rho_p=1000~kg~m^{-3}$ are injected
into the flow.
The motion of
particles is described by a set of ordinary differential
equations for particle velocity and position.
For particles much heavier than the fluid ($\rho_{p}/\rho \gg 1$) the most
significant forces are Stokes drag and buoyancy
and Basset force can be neglected being
an order of magnitude smaller \cite{et92}.
In the present simulations, the aim is to minimize the
number of degrees of freedom by keeping the simulation setting
as simplified as possible; thus
the effect of gravity has also been
neglected. 
Particles, which are assumed pointwise, rigid and spherical,
are injected into the flow at concentration low enough to neglect
particle collisions.
The effect of particles onto the turbulent field
is also neglected (one-way coupling assumption).

With the above assumptions, a
simplified version of the Basset-Boussinesq-Oseen
equation \cite{cst98} is obtained.
In vector form it writes:
\begin{eqnarray}
\label{partpos}
\frac{d{\bf x}_p}{dt} = {\bf v}~,
\end{eqnarray}
\vspace{-0.6cm}
\begin{eqnarray}
\label{partvel}
\frac{d{\bf v}}{dt} =
{\frac{\mathbf{u}_s -\mathbf v }{\tau_p} (1 + 0.15 Re_p^{0.687})}~,
\end{eqnarray}
where ${\bf x}_p$ is the particle position,
${\bf v}$ is the particle velocity, and
${\bf u}_s$ is the fluid velocity
at the particle position, ${\bf u}({\bf x}_p(t),t)$.
The righthand side of Eq. (\ref{partvel}) represent the drag force per unit mass acting of the particle, $D$, in which $\tau_p$ is the particle relaxation time, defined as $\tau_{p} = \rho_{p}d^{2}_{p}/18\mu$, $d_{p}$ being the particle diameter, and 
$Re_p = d_{p}|{\bf v} - {\bf u}|/\nu$ is the particle Reynolds number, $\nu$ being the fluid kinematic viscosity.

The fully-developed channel flow previously described has been simulated through DNS.
The fluid governing equations (omitted here for the sake of brevity) are discretized
using a pseudo-spectral method based on transforming the field variables into wavenumber space, using Fourier
representations for the periodic
streamwise and spanwise directions and a Chebyshev representation for the
wall-normal (non-homogeneous) direction.
A two level, explicit
Adams-Bashforth scheme for the non-linear terms, and an implicit
Crank-Nicolson method for the viscous terms are employed
for time advancement.
Further details of the method can be found in previous articles, e.g. \cite{PB_POF_96}.

The calculations were performed
on a computational domain of size
$4 \pi h \times 2 \pi h \times 2 h$
in $x$, $y$ and $z$ respectively.
The size of the computational domain in wall units
is $1885\times942\times300$. Wall
units are obtained combining $u_{\tau}$, $\nu$ and $\rho$. The computational domain is discretized in physical space
with $128\times128\times129$ grid points (corresponding to $128\times128$
Fourier modes and to 129 Chebyshev coefficients in the wavenumber space). 
This is the minimum number of grid points required in each direction
to ensure that the grid spacing is always smaller than the smallest
flow scale and that the limitations imposed
by the point-particle approach are satisfied \cite{Marchioli2008}. 

To calculate particle trajectories in the DNS flow
field, $6^{th}$-order
Lagrangian polynomials are used to interpolate fluid velocities at
the particle position. With this velocity, Eqns. (\ref{partpos}) and (\ref{partvel}) are advanced in time
using a $4^{th}$-order Runge-Kutta scheme.

At the beginning of the simulation, particles
are distributed homogeneously over
the computational domain and their initial
velocity is set equal to that of the fluid
at the particle initial position. Periodic boundary conditions are imposed on particles
moving outside the computational domain in the
homogeneous directions,
and perfectly-elastic collisions are assumed at the smooth walls
when the particle center is at a distance lower than one
particle radius from the wall.
For the simulations presented here,
large samples of $10^{5}$ particles, characterized by
different response times, are considered. When
the particle response time is made
dimensionless using wall variables,
the Stokes number for each particle
set is obtained as $St=\tau_p^+=\tau_p/\tau_f$
where $\tau_f=\nu/u_{\tau}^2$ is the
viscous timescale of the flow.
Table~\ref{table:part} shows all the parameters
of the particles injected into the flow field.
\begin{table}[t]
\caption{\label{table:part} Particle parameters; $\tau_p$ is the particle relaxation time, $d_p$ the particle diameter, $V_s$ the particle settling velocity and $Re_p$ the particle Reynolds number.}
\begin{small}
\begin{center}
\begin{tabular}{c c c c c c}
\br
$St$ & $\tau_{p}~(s)$ & $d_{p}^{+}$ & $d_{p}$ (${\mu}m$) & $V_{s}^{+}=g^+\cdot St$  & $Re_{p}^{+}=V_s^+ \cdot d_{p}^{+} / \nu^+$\\
\mr
1   & $1.133 \cdot 10^{-3}$ & $0.153$ & $~20.4$ & $0.0943$ & $0.01443$\\
5   & $5.660 \cdot 10^{-3}$ & $0.342$ & $~45.6$ & $0.4717$ & $0.16132$\\
25  & $28.32 \cdot 10^{-3}$ & $0.765$ & $102.0$ & $2.3584$ & $1.80418$\\
\br
\end{tabular}
\vspace{0.3cm}
\end{center}
\end{small}
\end{table}

The non-dimensional timestep size used for particle tracking was
chosen to be equal to the non-dimensional timestep size used for
the fluid, $\delta t^+ = 0.045$; the total tracking time in DNS 
was, for each particle set, $t^+ = 21150$ (in wall units).
\subsection{A-priori LES methodology and computation of the ideal correction}
\label{IC}
In the {\em a-priori} tests the Lagrangian tracking of particles is
carried out starting from the filtered fluid velocity field,
$\bar{{\bf u}}$,
obtained through explicit filtering of the DNS velocity by means
of either a cut-off or a top-hat filter. Both filters are applied
in the homogeneous streamwise and spanwise directions in the
wave number space:
\begin{eqnarray}
\label{cut-off}
\bar{u}_i ({\bf x},t) = FT^{-1} \left\{
\begin{array}{ll}
G(\kappa_1) \cdot G(\kappa_2) \cdot
\hat{u}_i (\kappa_1,\kappa_2,z,t)~~~~~~{\text{if}}~~| \kappa_j | \le |\kappa_c|
~~{\text{with}}~~j=1,2~,\\
0 \hspace{5.75cm} \text{otherwise~.}
\end{array}
\right.
\end{eqnarray}
where $FT$ is the 2D Fourier Transform, $\kappa_c= \pi / \Delta$ is the
cutoff wave number ($\Delta$ being the filter width in the physical
space), $\hat{u}_i (\kappa_1,\kappa_2,z,t)$
is the Fourier transform of the $i$th component of the fluid velocity, namely
$\hat{u}_i (\kappa_1,\kappa_2,z,t)=FT [{u}_i ({\bf x},t) ]$ and $G(\kappa_i)$ is
the filter transfer function:
\begin{eqnarray}
\label{transfer-function}
G(\kappa_j) = \left\{
\begin{array}{ll}
1~~~~~~~~~~~~~~~~~~~{\text{for the cut-off filter~,}}\\
\frac{sin(\kappa_j \Delta/2)}{\kappa_j \Delta/2} ~~~~~~~~~\text{for the top-hat filter~.}
\end{array}
\right.
\end{eqnarray}
Three different filter widths have been considered,
corresponding to a grid Coarsening Factor (CF) in each homogeneous direction of 2, 4 and 8 with respect to DNS, i.e. to $64 \times 64$, $32 \times 32$ and $16 \times 16$ Fourier modes in the homogeneous directions respectively. Note that a coarsening factor of 8 leads to a very coarse resolution and is considered as an extreme case of under-resolved LES, while CF of 2 and 4 correspond to resolutions more currently used in LES. In the wall-normal direction data are not filtered,
since often in LES the wall-normal resolution is DNS-like. \cite{pope-faq}

In a-priori tests, the filtered flow velocity $\bar{\bf u}$ is used in Eq. (\ref{partvel}) for Lagrangian particle tracking, instead of the DNS one. The SGS correction term is computed by imposing that the particle trajectories tracked in a-priori LES coincide with those in DNS, at each time step, $t^n$, and for each particle, $k$, we impose:
\begin{eqnarray}
\label{cond_corr}
\mathbf{x}^{LES}_k(t^n)=\mathbf{x}^{DNS}_k(t^n) \; \; ; \; \; \mathbf{v}^{LES}_k(t^n)=\mathbf{v}^{DNS}_k(t^n)
\end{eqnarray}
in which the superscript $LES$ denotes the particle position and velocity obtained in a-priori LES, while the superscript $DNS$ those obtained in DNS. The correction term computed for the particle $k$ at time $t^n$ can be expressed as follows:
\begin{eqnarray}
\label{corr_term}
\mathbf{\delta u}(\mathbf{x}_k(t^n),t^n) =   \mathbf{u}(\mathbf{x}_k(t^n),t^n) - \bar{\mathbf{u}}(\mathbf{x}_k(t^n),t^n)
\end{eqnarray}
in which $\mathbf{x}_k(t^n)$ is the position of particle $k$ at time $t^n$ and the superscripts $LES$ and $DNS$ have been dropped since the particle positions are forced to be the same in DNS and a priori LES. It is thus evident from Eq. (\ref{corr_term}) that in this way the correction term expresses the filtering error on the fluid velocity seen by the particles, while the error accumulation due to the progressive divergence of the trajectories computed in LES and DNS is eliminated by assuming the coincidence of the trajectories. As previously discussed in the Introduction, this is obviously an idealized situation, but it allows a contribution to the error in Lagrangian tracking of particles in LES, i.e. the one due to filtering of the fluid velocity, to be singled out and characterized. Note also that Eq. (\ref{corr_term}) gives the difference between the DNS and the filtered fluid velocities \textit{at the particle position}, i.e. computed by following the particle trajectories. As it will be shown also in the following, this may be substantially different from the Eulerian measure of the filtering error, classically computed in the literature, i.e. the difference between DNS and LES velocities at fixed points in space. 

Following Eq. (\ref{corr_term}), the SGS correction term to the fluid velocity seen by the particles has been computed for each particle along the trajectories computed in the DNS fluid velocity fields, i.e. at ${x}^{DNS}_k(t^n)$. The computation has been carried out for the three considered particle sets characterized by different inertia and for the different filter widths and types, for a total time $\Delta t^+ \simeq 8285$ and sampled every $\delta t^+=4.5$. It has been checked that this leads to converged statistics of $\mathbf{\delta u}$.
\section{Statistical moments of the SGS correction}
\label{statistics}

The Eulerian statistical moments of the correction term (Eq. (\ref{corr_term})) are presented and analyzed in this section. They are computed by dividing the computational domain in slabs having dimensions $L_x$, $L_y$ and $\Delta_z(i)$ ($i=1, N_z$), $\Delta_z(i)$ being the difference between two adjacent Chebyshev collocation points and $N_z$ the number of collocation points used in DNS. Averaging is carried out over the particles laying at a given time instant inside each slab and in time over the time interval specified in Sec. \ref{IC}. All the quantities shown herein are non dimensional, in wall units.
\subsection{Effects of particle inertia}
\label{statistics_inertia}
In order to analyze first the effects of particle inertia, let us fix the filter width and type. The influence of the filter will be investigated in Sec. \ref{statistics_filter}. We consider, in particular, the cut-off filter with a coarsening factor of 4 with respect to DNS.

\begin{figure}
\begin{minipage}[t]{18pc}
\vspace{0pt}
\includegraphics[width=18pc,angle=0.]{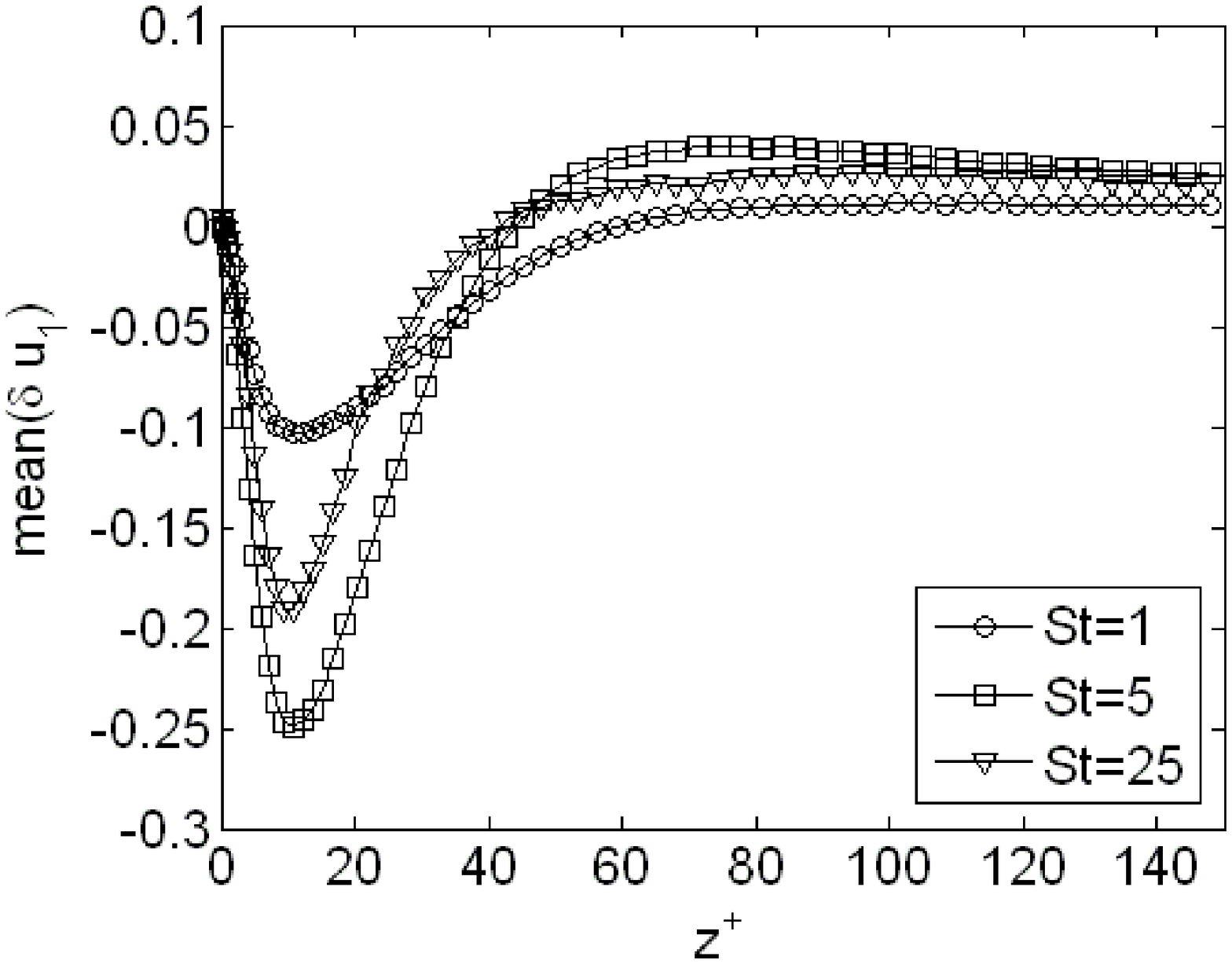}
\caption{\label{mean_IC_inertia_x} Mean values of the SGS velocity correction component in the streamwise direction as a function of $z^+$. Cut-off filter with CF=4.}
\end{minipage}\hspace{2pc}
\begin{minipage}[t]{18pc}
\vspace{0pt}
\includegraphics[width=18pc,angle=0.]{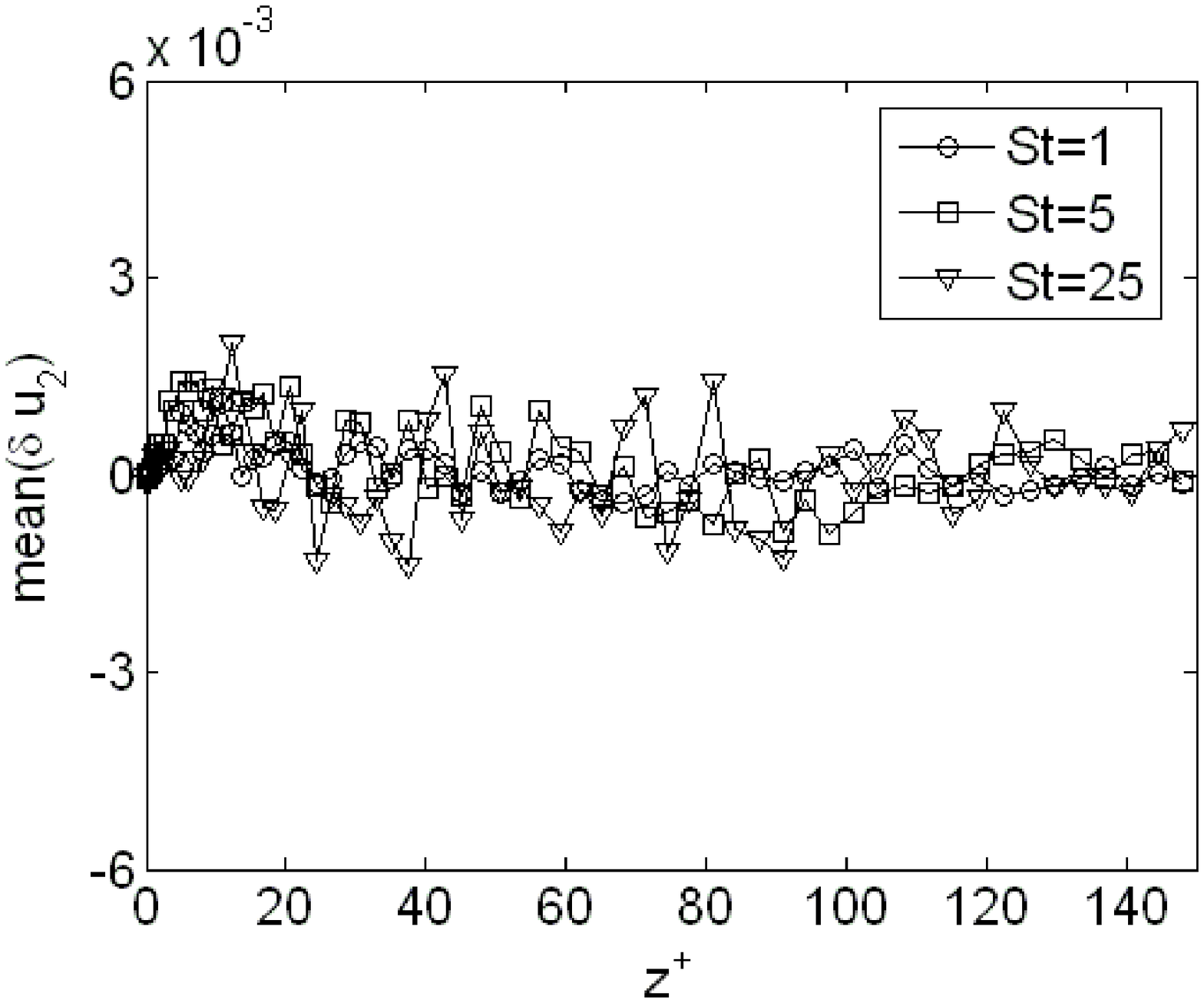}
\caption{\label{mean_IC_inertia_y} Mean values of the SGS velocity correction component in the spanwise direction as a function of $z^+$. Cut-off filter with CF=4.}
\end{minipage}\hspace{2pc}
\begin{minipage}[t]{18pc}
\vspace{0pt}
\includegraphics[width=18pc,angle=0.]{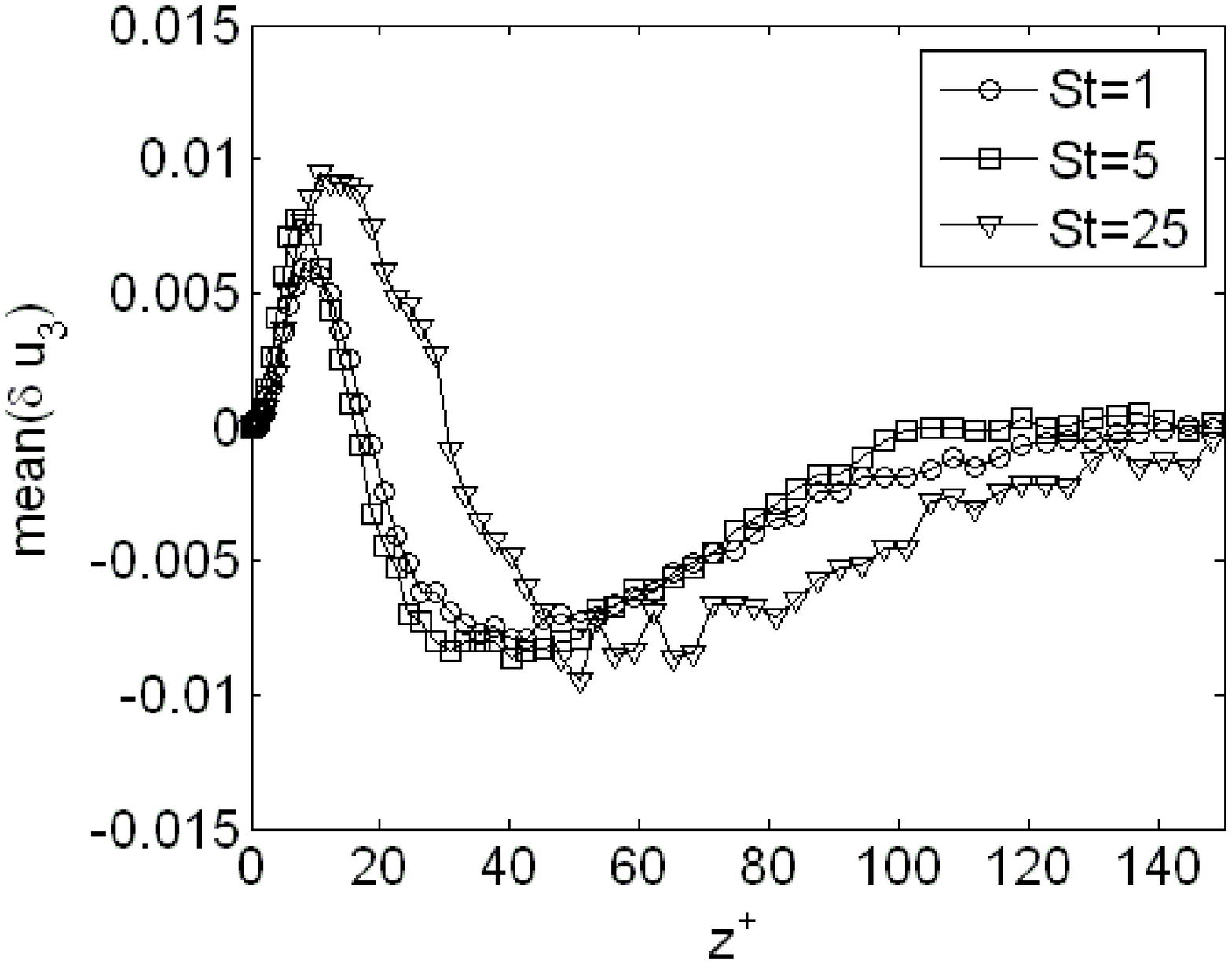}
\caption{\label{mean_IC_inertia_z} Mean values of the SGS velocity correction component in the wall-normal direction as a function of $z^+$. Cut-off filter with CF=4.}
\end{minipage}\hspace{2pc}
\begin{minipage}[t]{18pc}
\vspace{0pt}
\includegraphics[width=18pc,angle=0.]{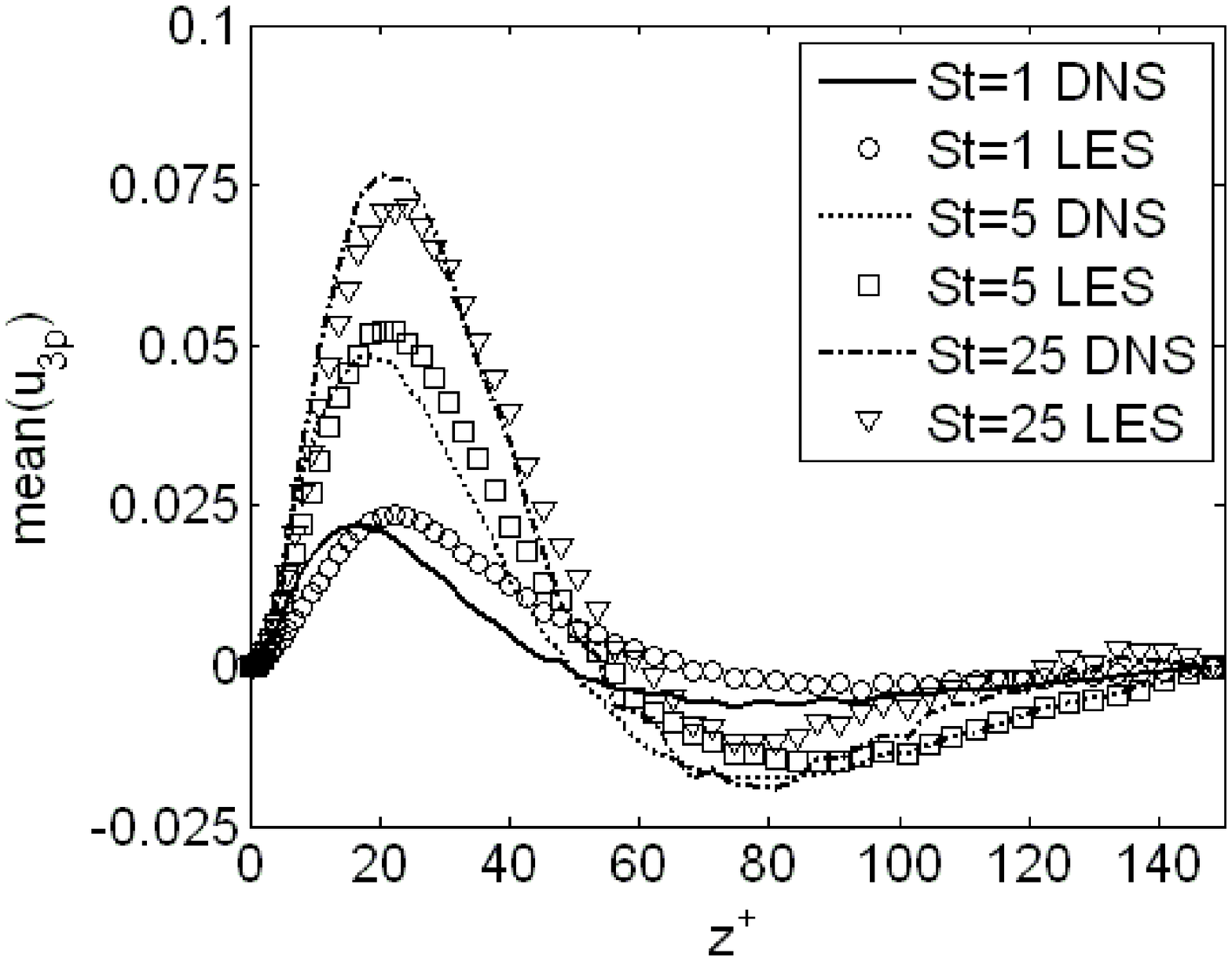}
\caption{\label{fluidvel_part} Effects of filtering on the fluid mean normal velocity seen by the particles. Cut-off filter with CF=4.}
\end{minipage}\hspace{2pc}
\end{figure}

Figures \ref{mean_IC_inertia_x}-\ref{mean_IC_inertia_z} show the mean value of the components of the SGS velocity correction term along the streamwise, spanwise and normal directions, computed for the considered filter and for different particle inertia, as a function of the distance from the wall. While the mean values of the spanwise component of the correction are very low along the whole channel, the spanwise and the normal components show significant (negative or positive) mean values in the near wall region, with peaks located approximately at $z^+ \simeq 15-20$. A first observation is that, if the correction term was computed at fixed grid points, instead of following the particle trajectories, its mean value would have been rigorously equal to zero in all the directions, since the filter does not affect the mean value of the velocity. Note that, in priori and a-posteriori tests in the literature, it is usual to analyze the effects of filtering (or, more in general, of errors) on the LES fluid velocity fields at fixed points (see e.g. \cite{kv05,Marchioli2008}). When dealing with Lagrangian tracking of particles, conversely, it seems to be more appropriate to study the effects of filtering on the fluid velocity seen by the particles. As it will be shown, the behavior of the mean streamwise and normal components of $\mathbf{\delta u}$ is indeed related to the effects of filtering on the turbulent structures, and in particular on near-wall structures, combined with the preferential sampling of particles. 

Let us start by analyzing the streamwise component of $\mathbf{\delta u}$. As well known, the near wall region is dominated by the presence of low- and high-speed streaks and it has been shown \cite{Picciotto2005,Picciotto2005b,Soldati2009} that inertial particles tend to preferentially sample low-speed streaks. The effect of filtering is to attenuate the velocity fluctuations and therefore to smooth the streaks. As a consequence, the correction term in mean tends to reintroduce the effect of low-speed streaks smoothened by filtering and, thus, in mean it tends to decrease the LES fluid velocity seen by the particles near the wall (negative peaks of mean $\delta u_1$ in the near wall region). As for the effects of inertia, the behavior of the mean streamwise component of the correction term is qualitatively the one previously described for all the considered particle sets. Quantitatively, the mean correction is more important for the particles having larger inertia (St=5 and 25). This is consistent with the previous explanation of the origin of the mean values; indeed, for the considered Reynolds number, the particles characterized by St=5 and 25 have been found in previous DNS studies \cite{Picciotto2005,Picciotto2005b,Soldati2009} to preferentially concentrate more than particles of smaller inertia. 

As for the mean wall normal component (Fig. \ref{mean_IC_inertia_z}), positive values of the SGS correction term are found in the near wall region, corresponding to a correction oriented away from the wall, while it becomes negative as $z^+$ increases, to eventually vanish moving towards the center of the channel. This behavior is again related to the effects of filtering on the normal fluid velocity seen by the particles. It has been observed in DNS that the mean fluid velocity seen by the particles is positive in the near wall region \cite{Picciotto2005}, due to the fact that in the near-wall region inertial particles preferentially sample ejection-like environments, in spite they have a mean drift directed towards the wall. Fig. \ref{fluidvel_part} reports the mean normal velocity of the fluid seen by the different considered sets of particles in DNS, confirming the previously described behavior. Note how again the preferential sampling of the ejection-like environments is more pronounced for particles having larger inertia. The same velocity profiles as for DNS obtained in a-priori LES are also shown in Fig. \ref{fluidvel_part}. The effect of filtering is to reduce the mean fluid normal velocity seen by the particles near the wall and to overestimate it moving towards the center of the channel. This is due to a smoothing of the near-wall structures, which, as for the streamiwse streaks, tends to attenuate the sweep events (see e.g. Fig. 6 in \cite{Marchioli2011}). In practice, the mean correction term is given by the difference between the mean wall-normal fluid velocities seen by the particles in DNS and in LES. Thus, the correction term in mean tends to be positive near the wall, with a peak at $z^+=15-20$, and eventually becomes negative by moving towards the channel center. There is a significant effect of particle inertia on the values of $z^+$ at which the positive and negative peaks are located and at which the correction changes sign, being larger for the particles having $St=25$ than for the other two considered particle sets. Note, however, that also the fluid velocity seen by the particles in DNS significantly varies with the particle inertia, this indicating a different interplay between the near-wall vortical structures, sweep and ejection events and particles, depending on particle inertia. Therefore, it is not surprising that the effects of filtering in wall-normal direction are also  different for different particle inertia. It will be shown in Sec. \ref{statistics_filter} that the mean correction term in the wall-normal direction is also significantly sensitive to the filter width.
 
Note that preferential concentration and the interaction with the near-wall turbulent structures have been identified in previous DNS studies
 \cite{Picciotto2005,Picciotto2005b,Soldati2009} as key ingredients to explain turbophoresis, i.e. the tendency of inertial particles to accumulate at solid walls. On the other hand, it has been observed \cite{kv05,k06,Marchioli2008} that LES underestimates the turbophoretic effect. The previous analysis of the mean SGS velocity correction term seems to indicate that this underestimation is related to the fact the near wall turbulent structures and their interactions with particles are incorrectly captured in LES due to errors introduced by filtering.

\begin{figure}
\begin{minipage}{18pc}
\includegraphics[width=18pc,angle=0.]{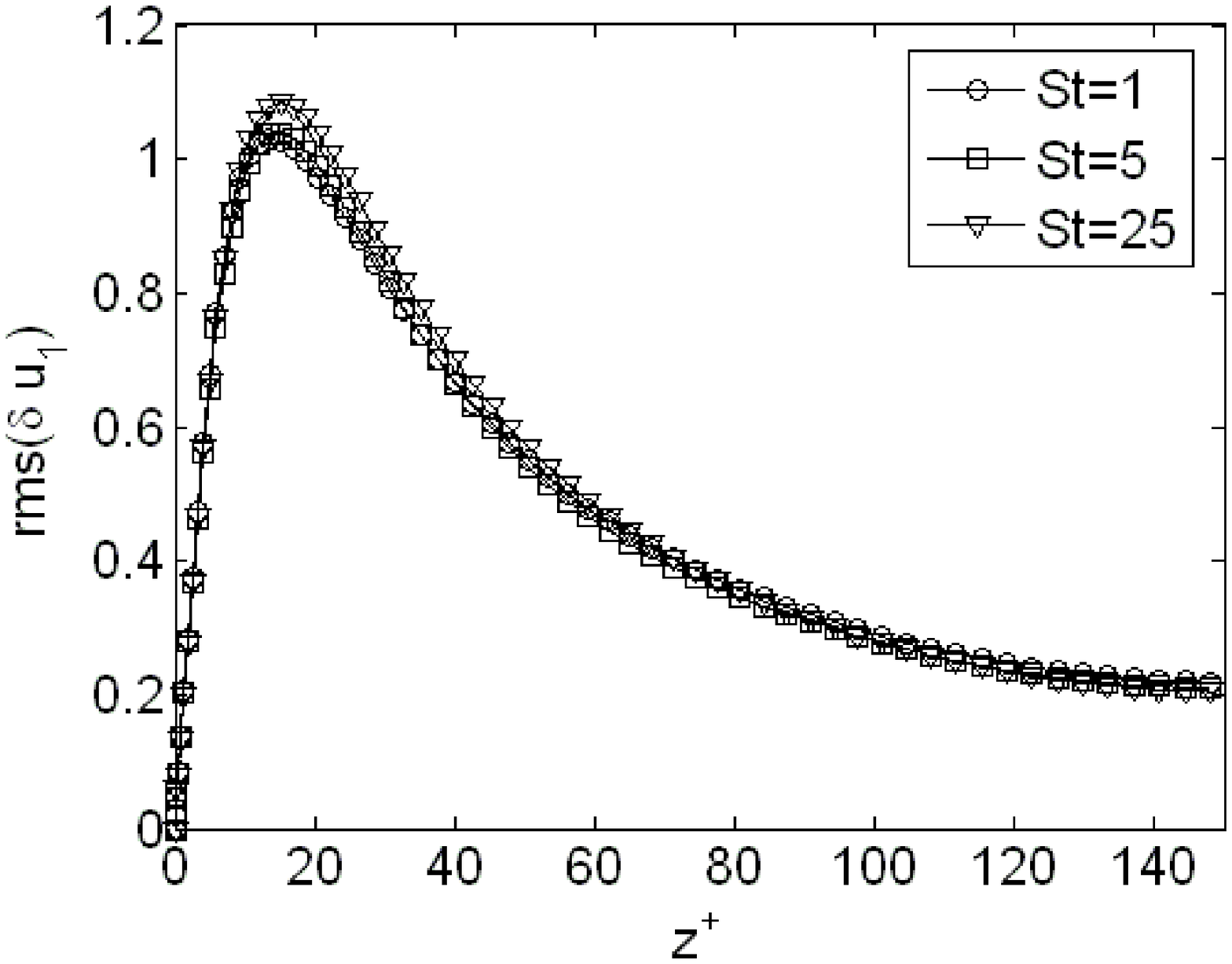}
\caption{\label{rms_IC_inertia_x} R.m.s values of the SGS velocity correction component in the streamwise direction as a function of $z^+$. Cut-off filter with CF=4.}
\end{minipage}\hspace{2pc}
\begin{minipage}{18pc}
\includegraphics[width=18pc,angle=0.]{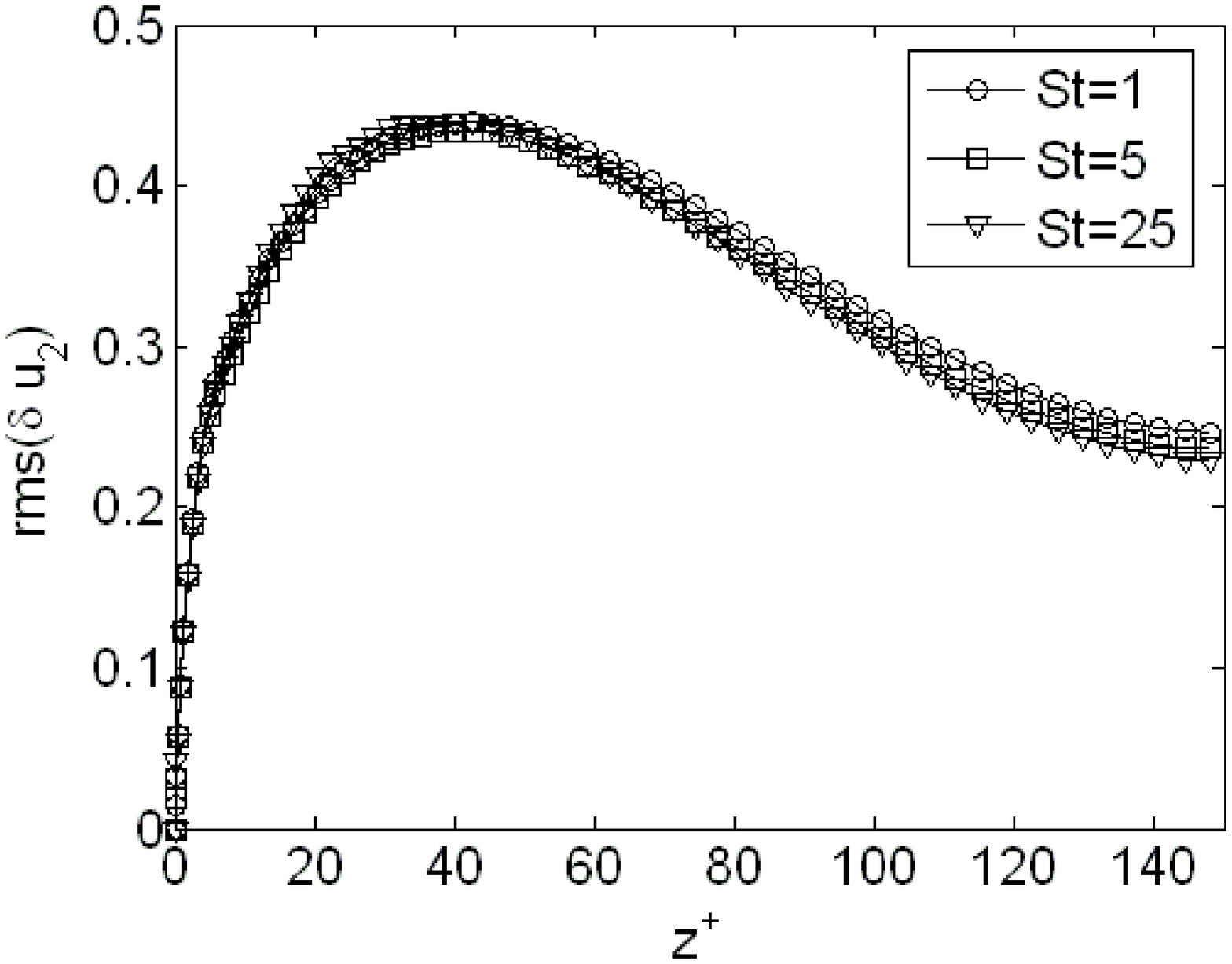}
\caption{\label{rms_IC_inertia_y} R.m.s values of the SGS velocity correction component in the spanwise direction as a function of $z^+$. Cut-off filter with CF=4.}
\end{minipage}\hspace{2pc}
\begin{minipage}[t]{18pc}
\vspace{0pt}
\includegraphics[width=18pc,angle=0.]{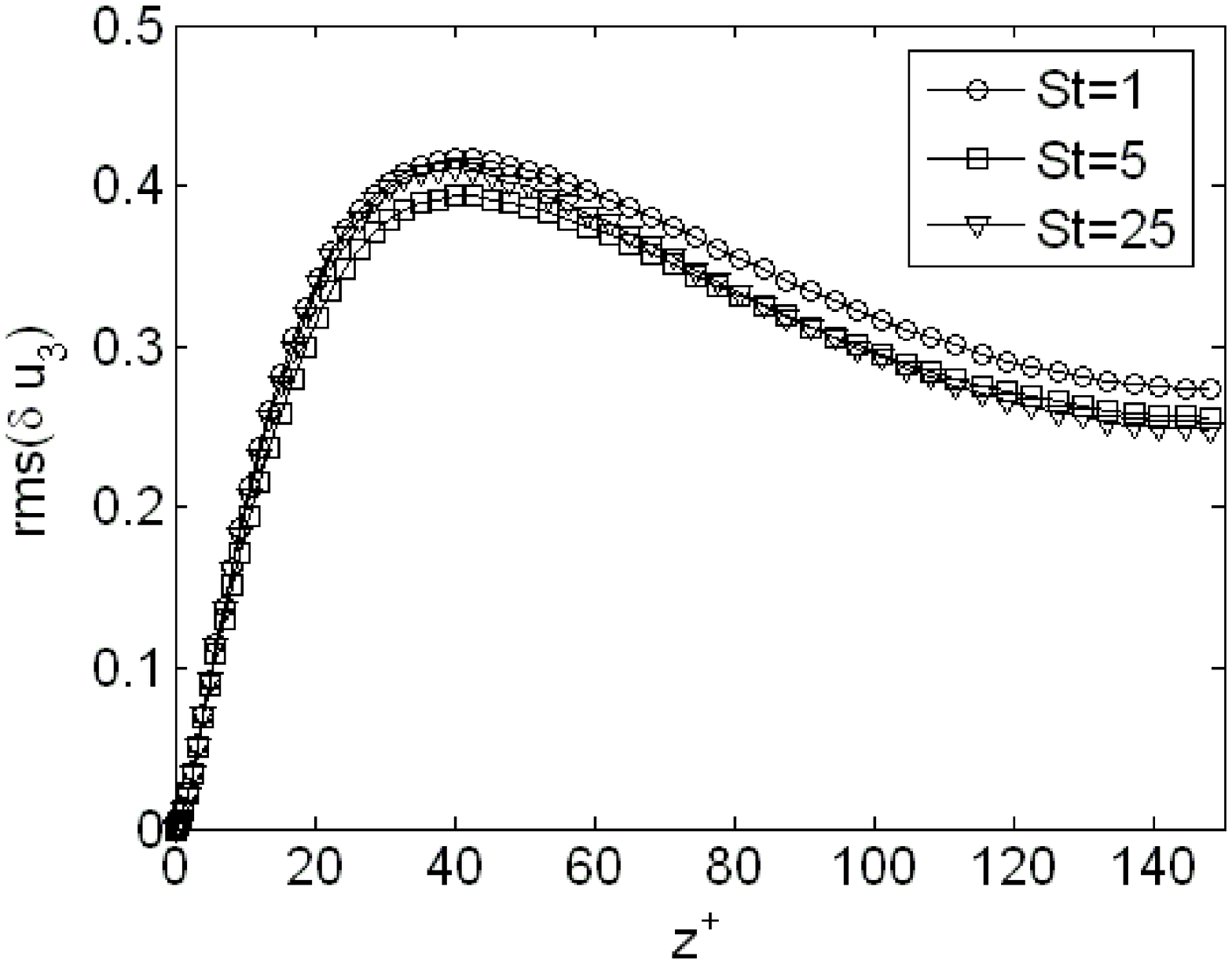}
\end{minipage}\hspace{2pc}
\begin{minipage}[t]{18pc}
\vspace{0pt}
\caption{\label{rms_IC_inertia_z} R.m.s values of the SGS velocity correction component in the wall-normal direction as a function of $z^+$. Cut-off filter with CF=4.}
\end{minipage}
\end{figure}

The r.m.s. profiles of the three components of the correction term are reported in Figs. \ref{rms_IC_inertia_x}-\ref{rms_IC_inertia_z}. In all directions, the main effect is to compensate the underestimation of fluid velocity fluctuations due to filtering. This reduction of the fluid velocity fluctuations in LES is well known since it can also be found in classical error analysis carried out at fixed points, and the SGS models previously used for LPT in LES fields, such as for instance approximate deconvolution or filtering inversion, are mainly aimed at counteracting this effect, i.e. at reintroducing the correct amount of fluid velocity fluctuations \cite{k06,Marchioli2008}. The r.m.s. profiles obtained for the three different sets of particles are very similar each other and thus no significant effects of particle inertia are present.

\begin{figure}
\begin{minipage}{18pc}
\includegraphics[width=18pc,angle=0.]{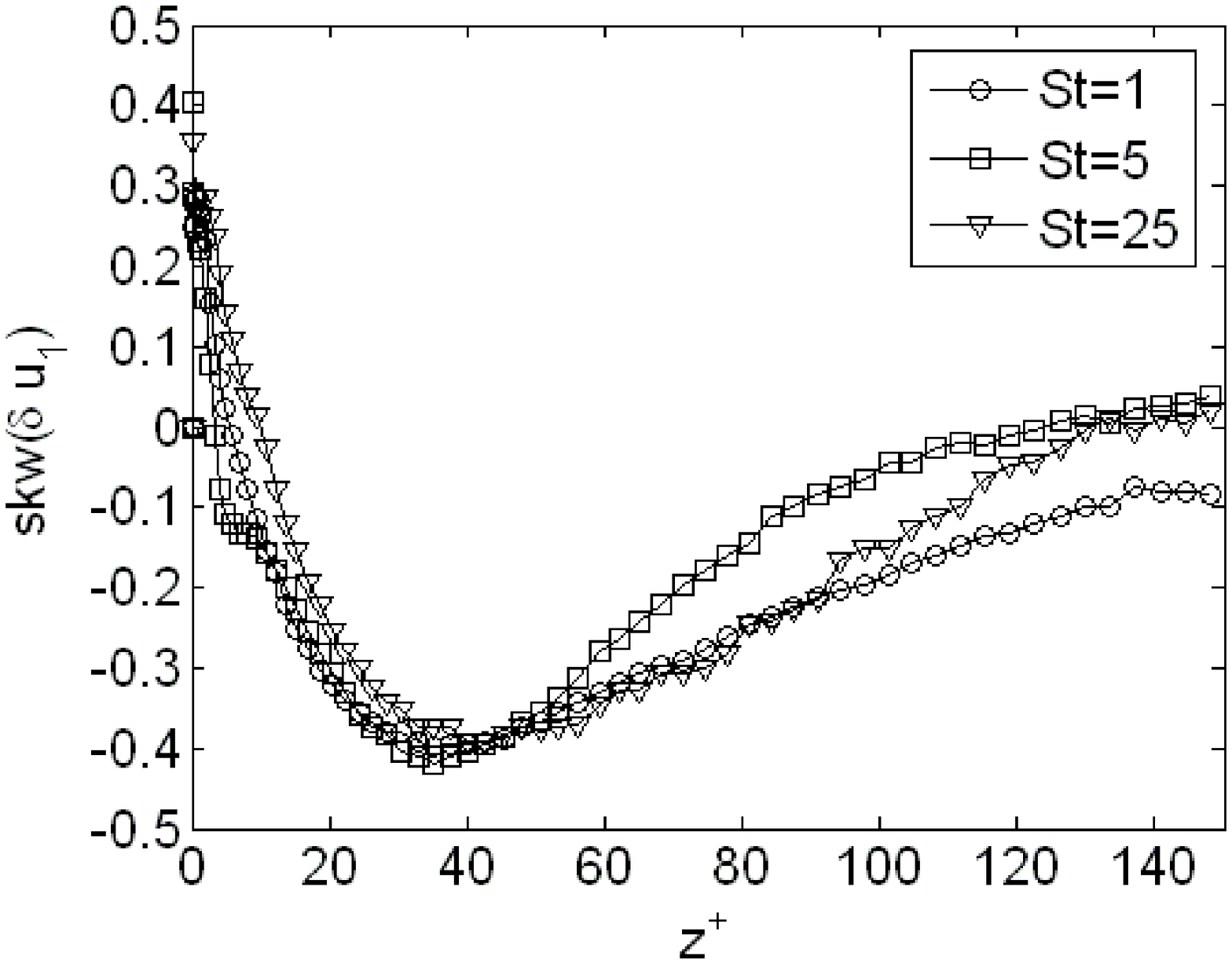}
\caption{\label{skew_IC_inertia_x} Skewness of the SGS velocity correction component in the streamwise direction as a function of $z^+$. Cut-off filter with CF=4.}
\end{minipage}\hspace{2pc}
\begin{minipage}{18pc}
\includegraphics[width=18pc,angle=0.]{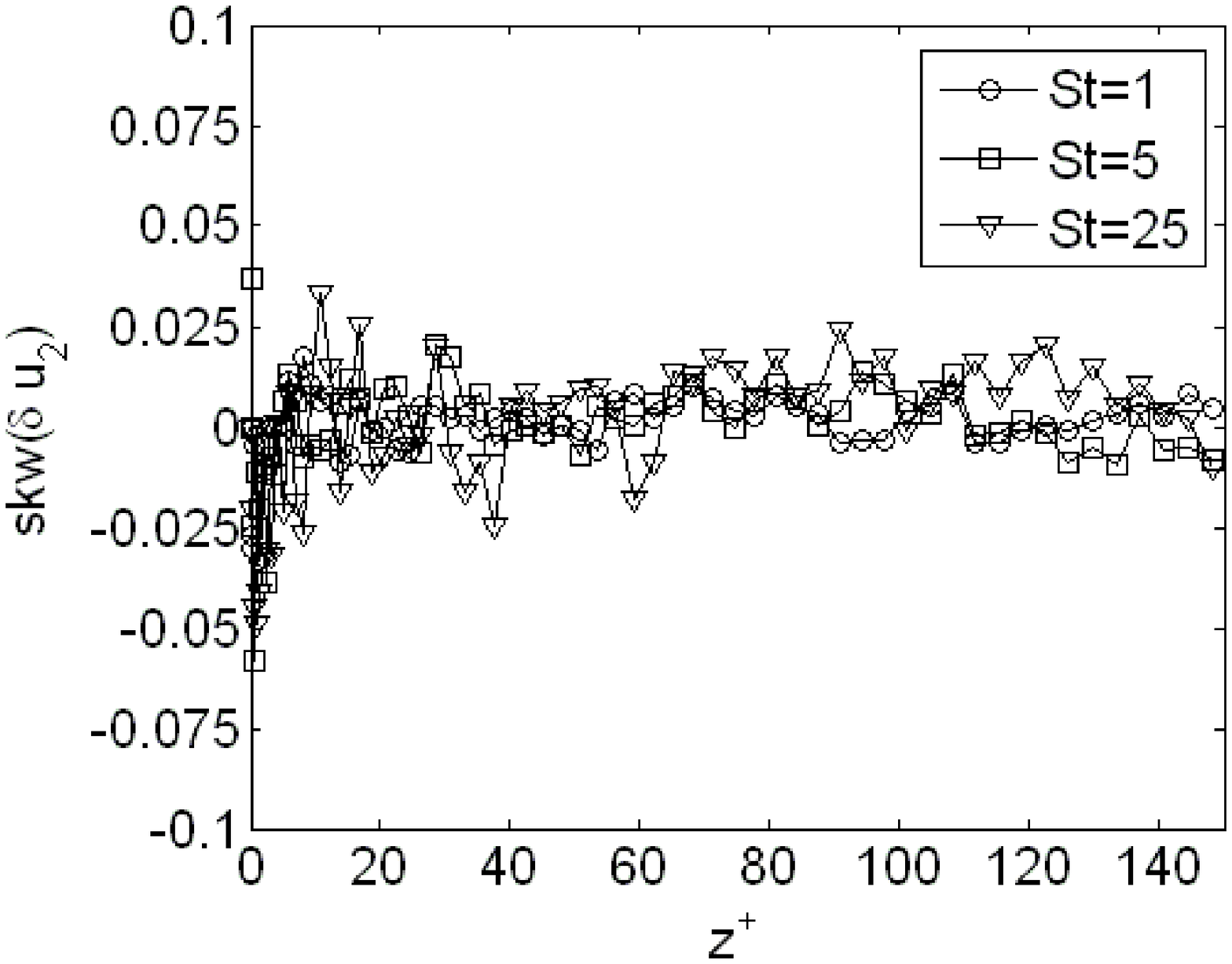}
\caption{\label{skew_IC_inertia_y} Skewness of the SGS velocity correction component in the spanwise direction as a function of $z^+$. Cut-off filter with CF=4.}
\end{minipage}\hspace{2pc}
\begin{minipage}[t]{18pc}
\vspace{0pt}
\includegraphics[width=18pc,angle=0.]{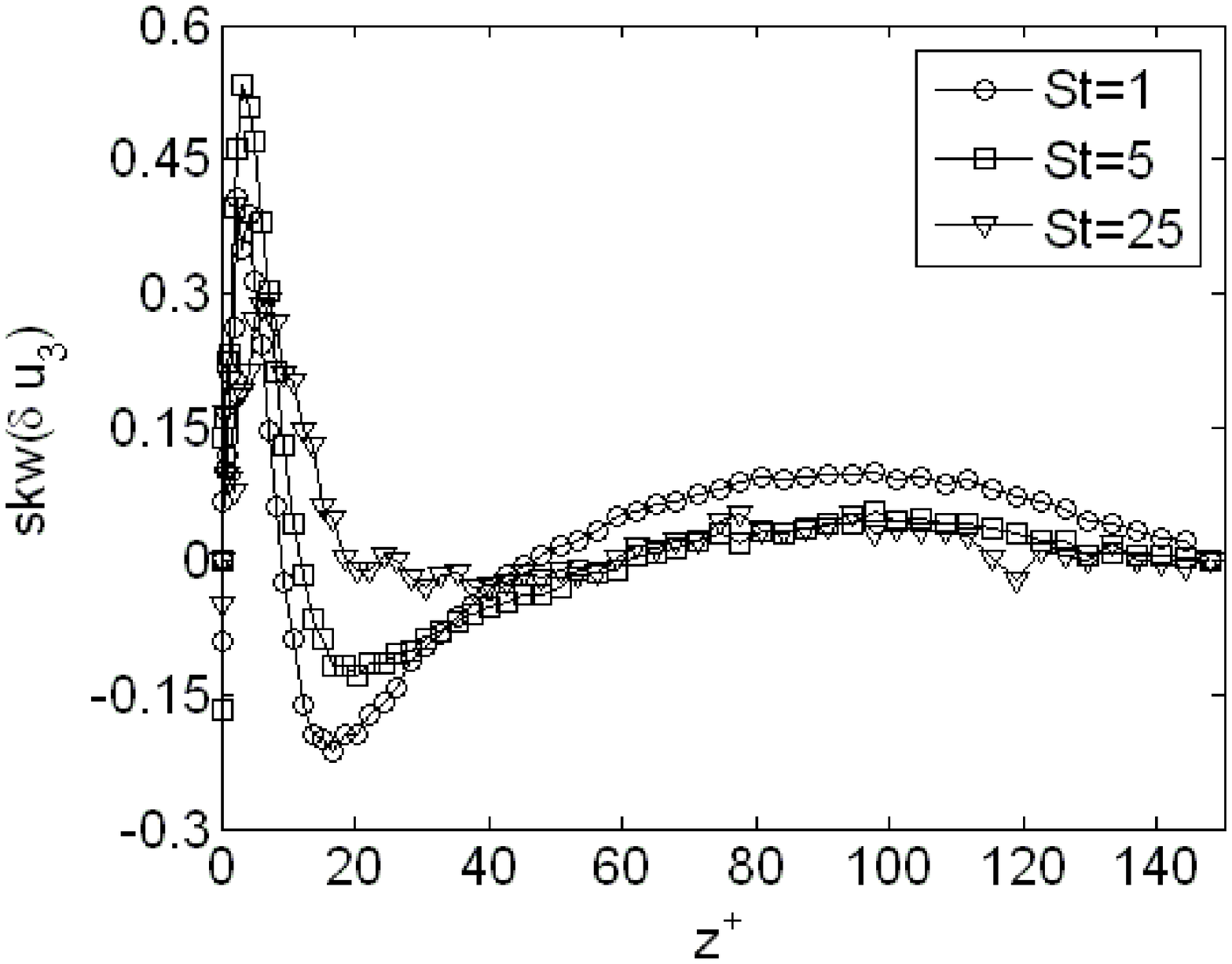}
\end{minipage}\hspace{2pc}
\begin{minipage}[t]{18pc}
\vspace{0pt}
\caption{\label{skew_IC_inertia_z} Skewness of the SGS velocity correction component in the wall-normal direction as a function of $z^+$. Cut-off filter with CF=4.}
\end{minipage}
\end{figure}

Figures \ref{skew_IC_inertia_x}-\ref{skew_IC_inertia_z} show the skewness profiles of the three components of the correction term. For the spanwise component, the skewness values remain very low along the whole channel, roughly oscillating around zero. Conversely, the streamwise component is characterized by significant values of the skewness, positive very close to the wall and negative in the largest part of the channel. Finally the skewness of the normal component of the correction is also characterized by positive significant values close to the wall, which become negative as $z^+$ increases, eventually become slightly positive again to vanish at the center of the channel. This behavior indicates that significant asymmetric deviations from Gaussianity are expected in the pdfs of the streamwise and normal components of the correction term, and, thus, that significant contributions to the streamwise and normal components of the correction term are given by isolated events of a given sign. This is consistent with the previous analysis, in which it has been shown that the particles mainly feel the effects of filtering on the low-speed streaks (negative velocity fluctuations) and on ejection events. The effects of particle inertia on the skewness of the streamwise component of the correction term appear to be more significant in the central part of the channel, while for the normal component larger differences are observed in the near wall region.

\begin{figure}
\begin{minipage}{18pc}
\includegraphics[width=18pc,angle=0.]{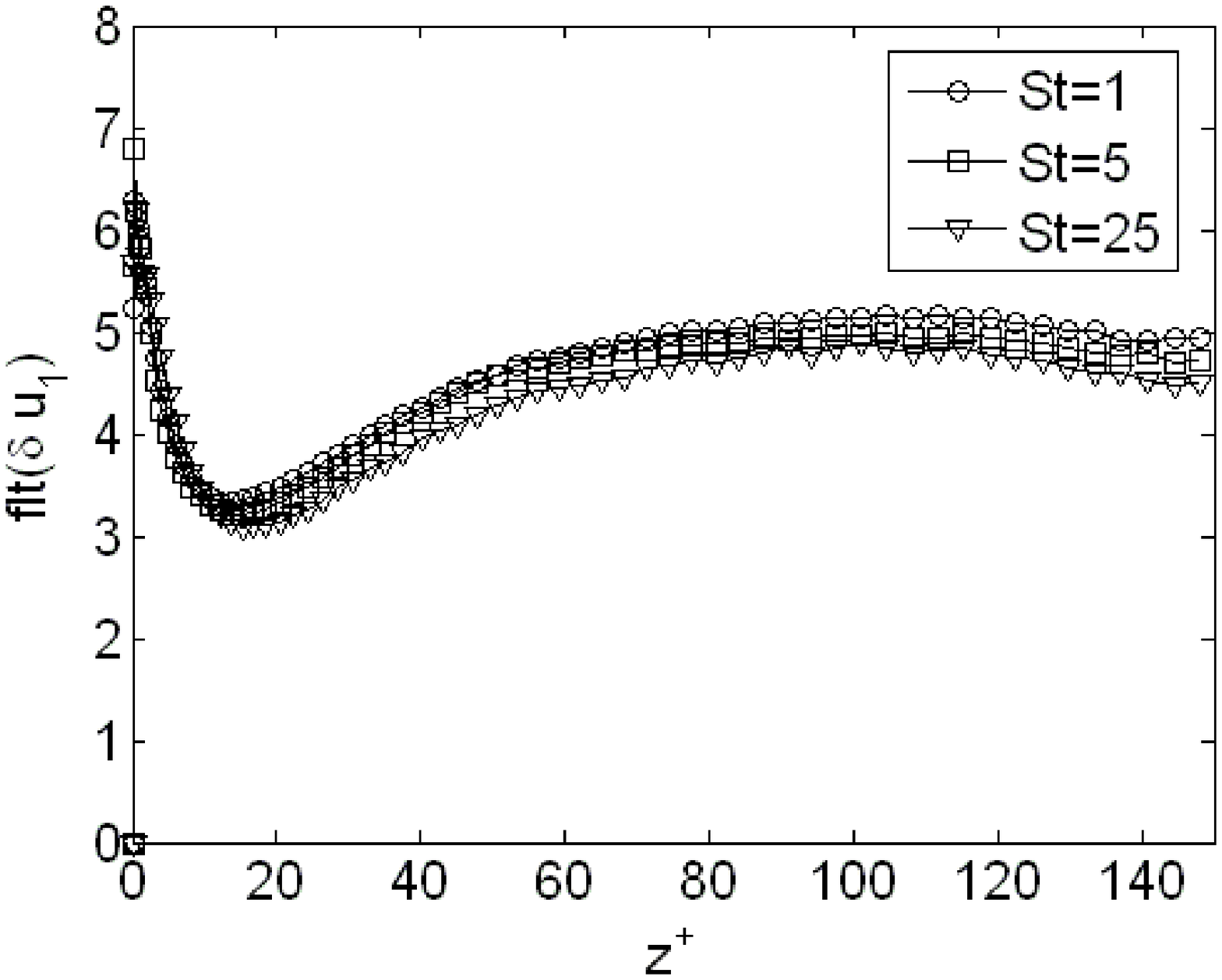}
\caption{\label{flat_IC_inertia_x} Flatness of the SGS velocity correction component in the streamwise direction as a function of $z^+$. Cut-off filter with CF=4.}
\end{minipage}\hspace{2pc}
\begin{minipage}{18pc}
\includegraphics[width=18pc,angle=0.]{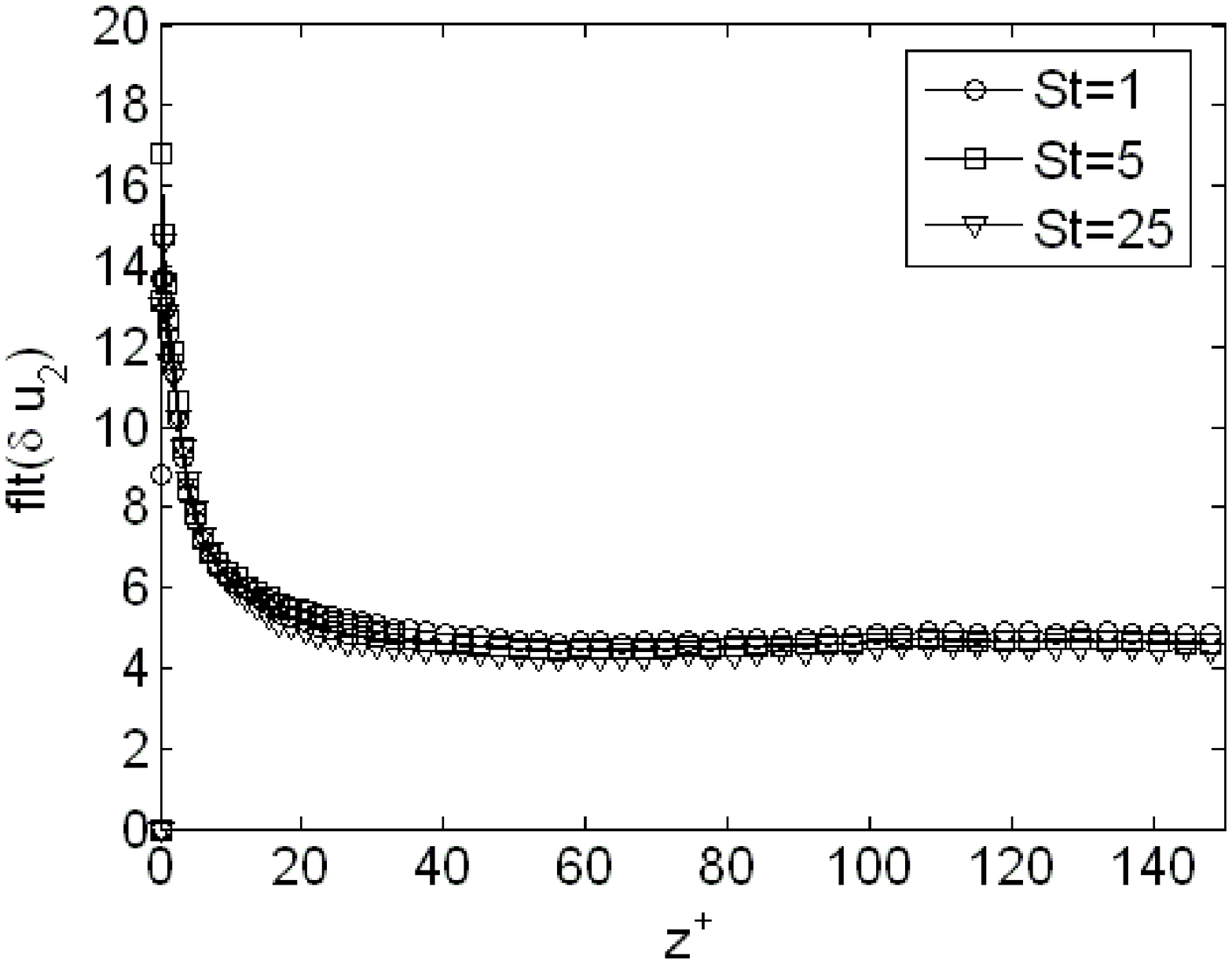}
\caption{\label{flat_IC_inertia_y} Flatness of the SGS velocity correction component in the spanwise direction as a function of $z^+$. Cut-off filter with CF=4.}
\end{minipage}\hspace{2pc}
\begin{minipage}[t]{18pc}
\vspace{0pt}
\includegraphics[width=18pc,angle=0.]{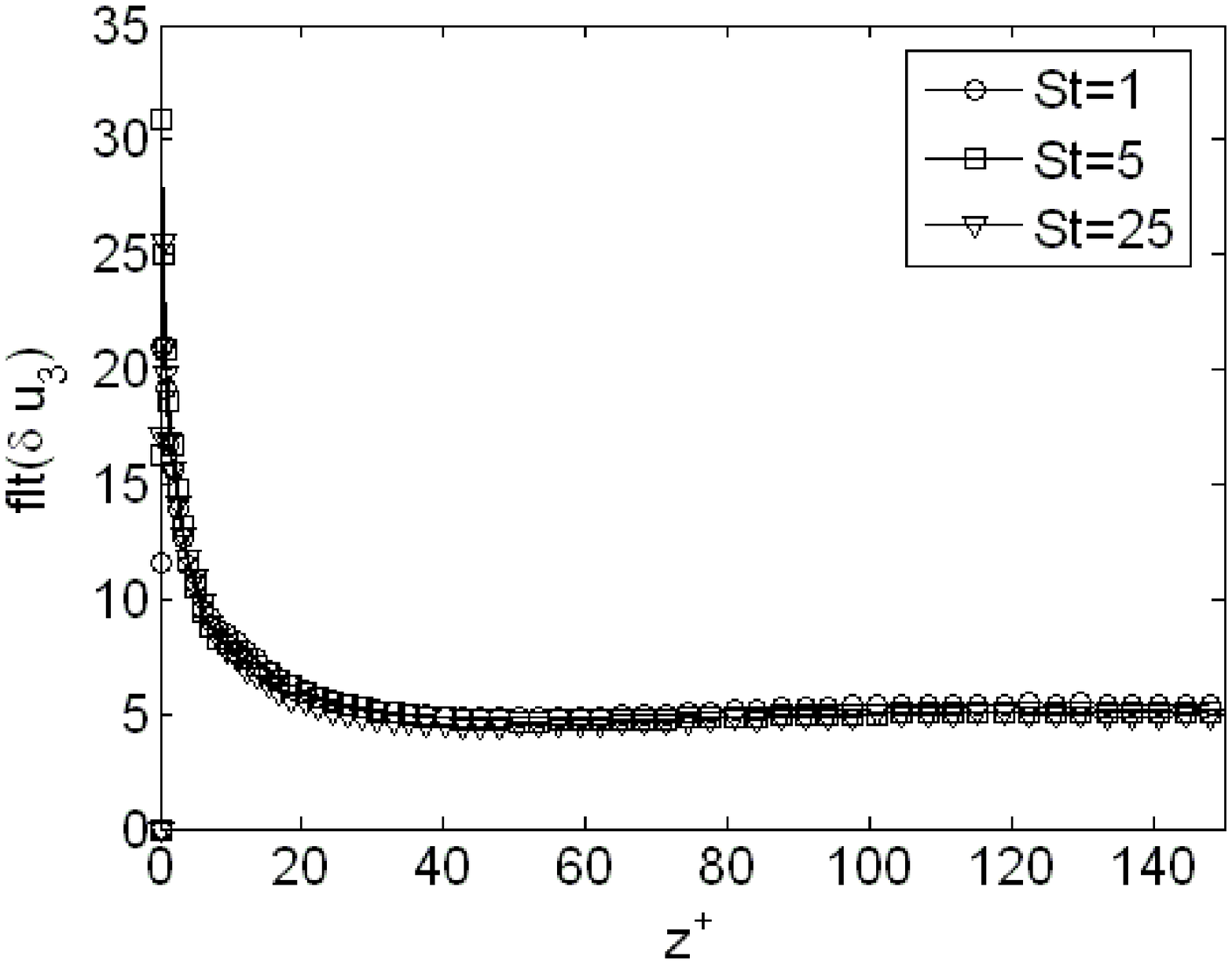}
\end{minipage}\hspace{2pc}
\begin{minipage}[t]{18pc}
\vspace{0pt}
\caption{\label{flat_IC_inertia_z} Flatness of the SGS velocity correction component in the wall-normal direction as a function of $z^+$. Cut-off filter with CF=4.}
\end{minipage}
\end{figure}

Finally, the flatness profiles of the three components of the correction term are reported in Figs. \ref{flat_IC_inertia_x}-\ref{flat_IC_inertia_z}. In all cases large values are present very near the wall decreasing to an almost constant value of about 4-5 moving towards the center of the channel (3 being the value corresponding to a Gaussian pdf). The effects of particle inertia are negligible. 

\subsection{Influence of filtering}
 \label{statistics_filter}
 
We investigate now the effect of filter type and width on the statistics of the SGS velocity correction and we consider the particle set characterized by St=5. 

\begin{figure}
\begin{minipage}{18pc}
\includegraphics[width=18pc,angle=0.]{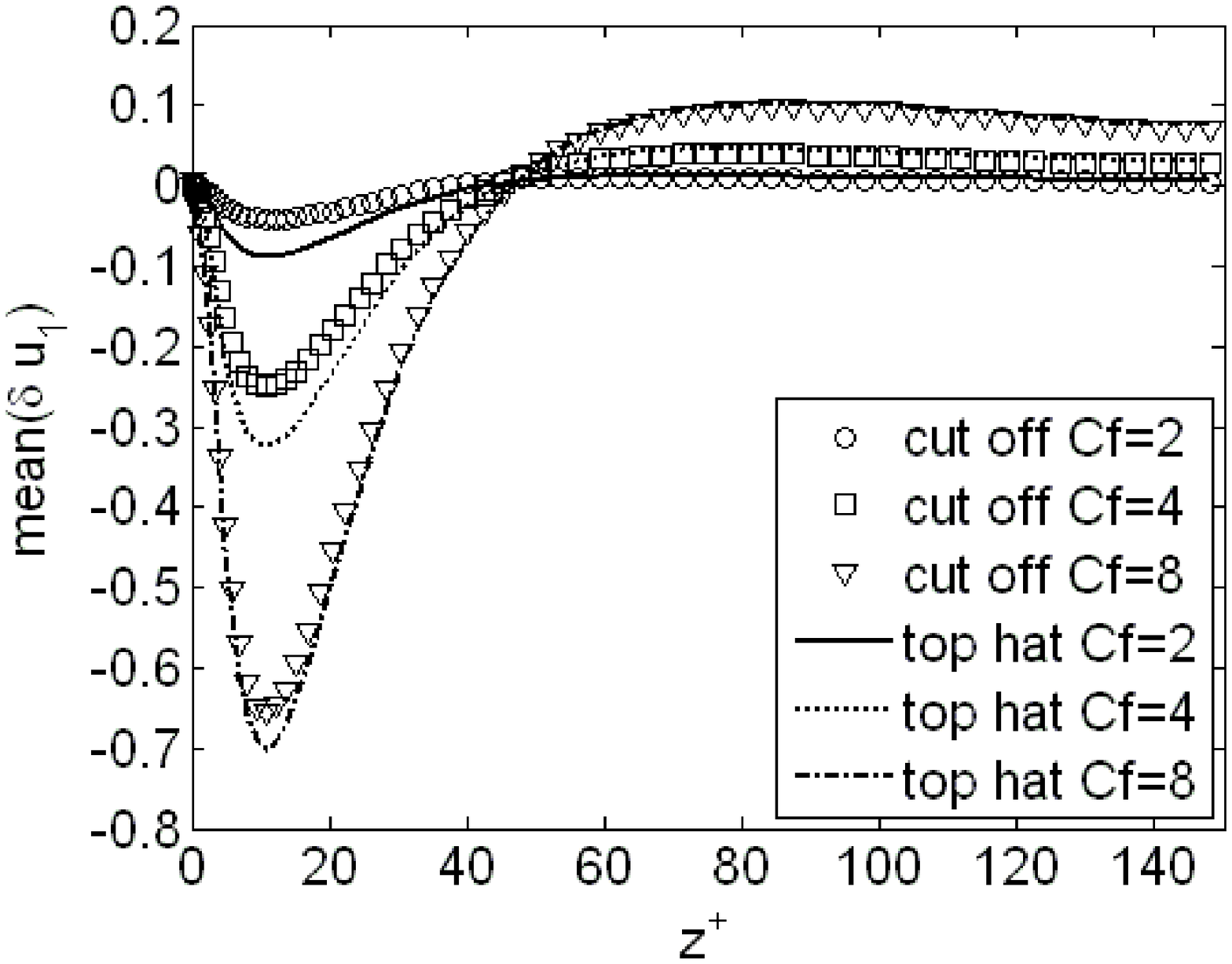}
\caption{\label{mean_IC_filter_x} Mean values of the correction term component in the streamwise direction as a function of $z^+$. St=5.}
\end{minipage}\hspace{2pc}
\begin{minipage}{18pc}
\includegraphics[width=18pc,angle=0.]{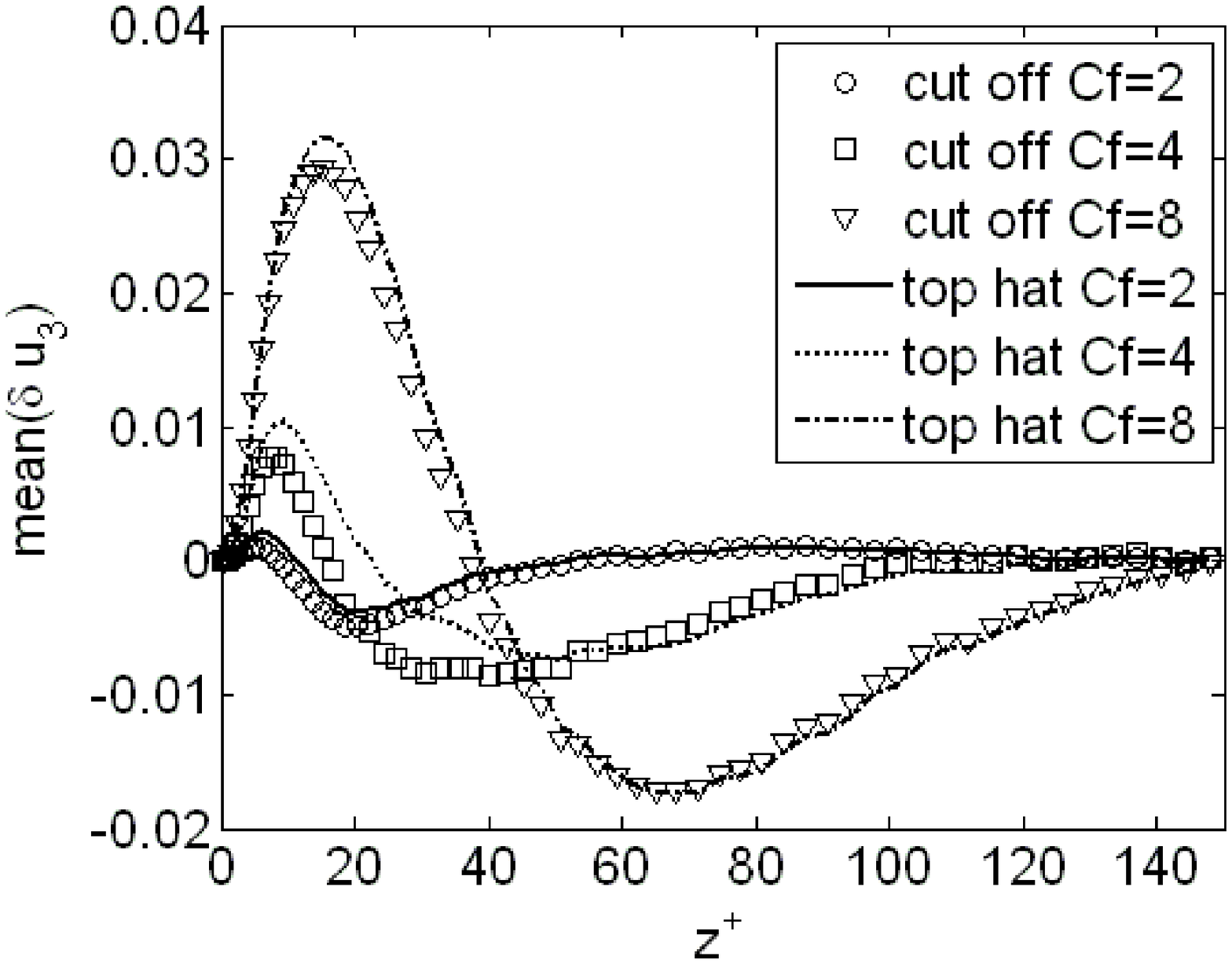}
\caption{\label{mean_IC_filter_z} Mean values of the correction term component in the wall-normal direction as a function of $z^+$. St=5.}
\end{minipage}\hspace{2pc}
\end{figure}

As for the mean value of the streamwise component of the correction, shown in Fig. \ref{mean_IC_filter_x}, the behavior is qualitatively similar for all the considered filters but the absolute value of the negative peak in the near wall region increases with the filter width. This could be expected, since, as pointed out in Sec. \ref{statistics_inertia}, this negative peak of the correction is related to the need of counteracting the attenuation of the low-speed streaks, preferentially sampled by particles, due to filtering. As for filter type, at fixed width, the top-hat filter introduces a stronger smoothing of the fluid velocity fluctuations than the cut-off one and thus the mean correction term is more important for the top-hat filter, especially for CF=2 and CF=4. The mean spanwise component of the correction term, for all the considered filters, always has very low values roughly oscillating around zero (not shown here for the sake of brevity). Figure \ref{mean_IC_filter_z} shows the mean normal component. Qualitatively, in all cases the behavior is the one described in Sec. \ref{statistics_inertia}, i.e. the mean correction is positive in the near-wall region, then becomes negative to eventually vanish at the center of the channel. However, the filter width not only affects the values of the positive and negative peaks, but also their location, which both tend to move away from the wall as the filter width increases. 

\begin{figure}
\begin{minipage}{18pc}
\includegraphics[width=18pc,angle=0.]{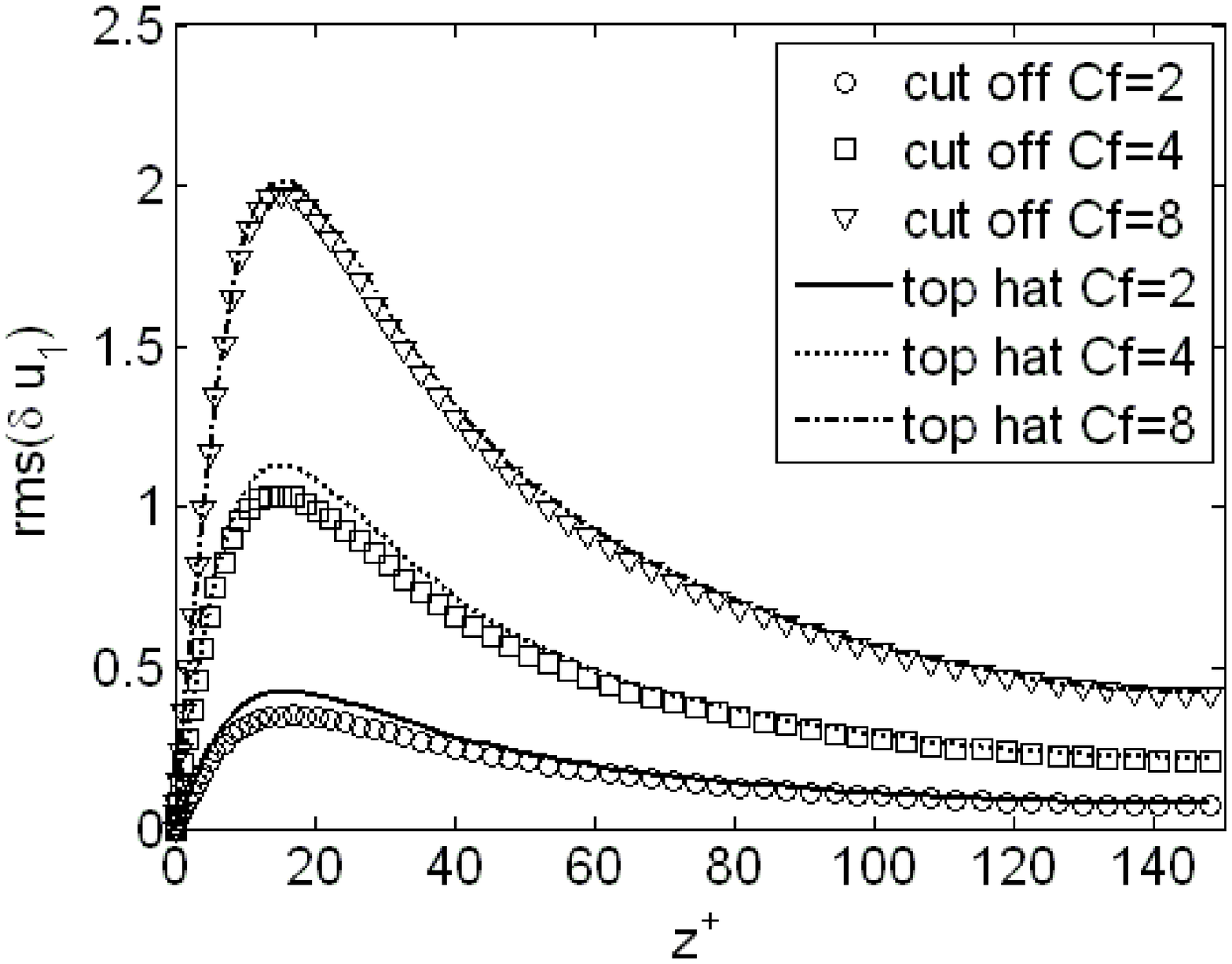}
\caption{\label{rms_IC_filter_x} R.m.s. values of the correction term component in the streamwise direction as a function of $z^+$. St=5.}
\end{minipage}\hspace{2pc}
\begin{minipage}{18pc}
\includegraphics[width=18pc,angle=0.]{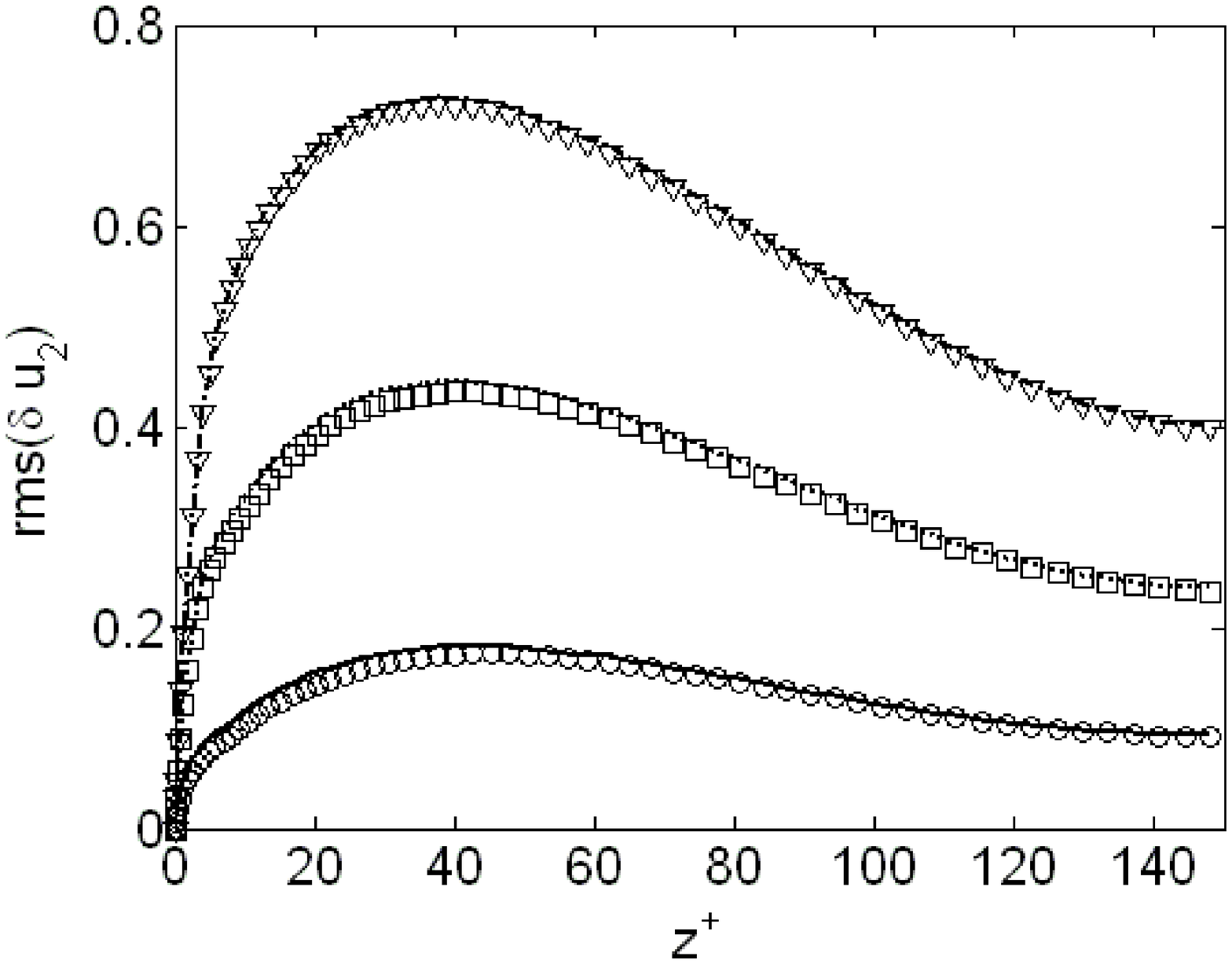}
\caption{\label{rms_IC_filter_y} R.m.s. values of the correction term component in the spanwise direction as a function of $z^+$. St=5.}
\end{minipage}\hspace{2pc}
\begin{minipage}[t]{18pc}
\vspace{0pt}
\includegraphics[width=18pc,angle=0.]{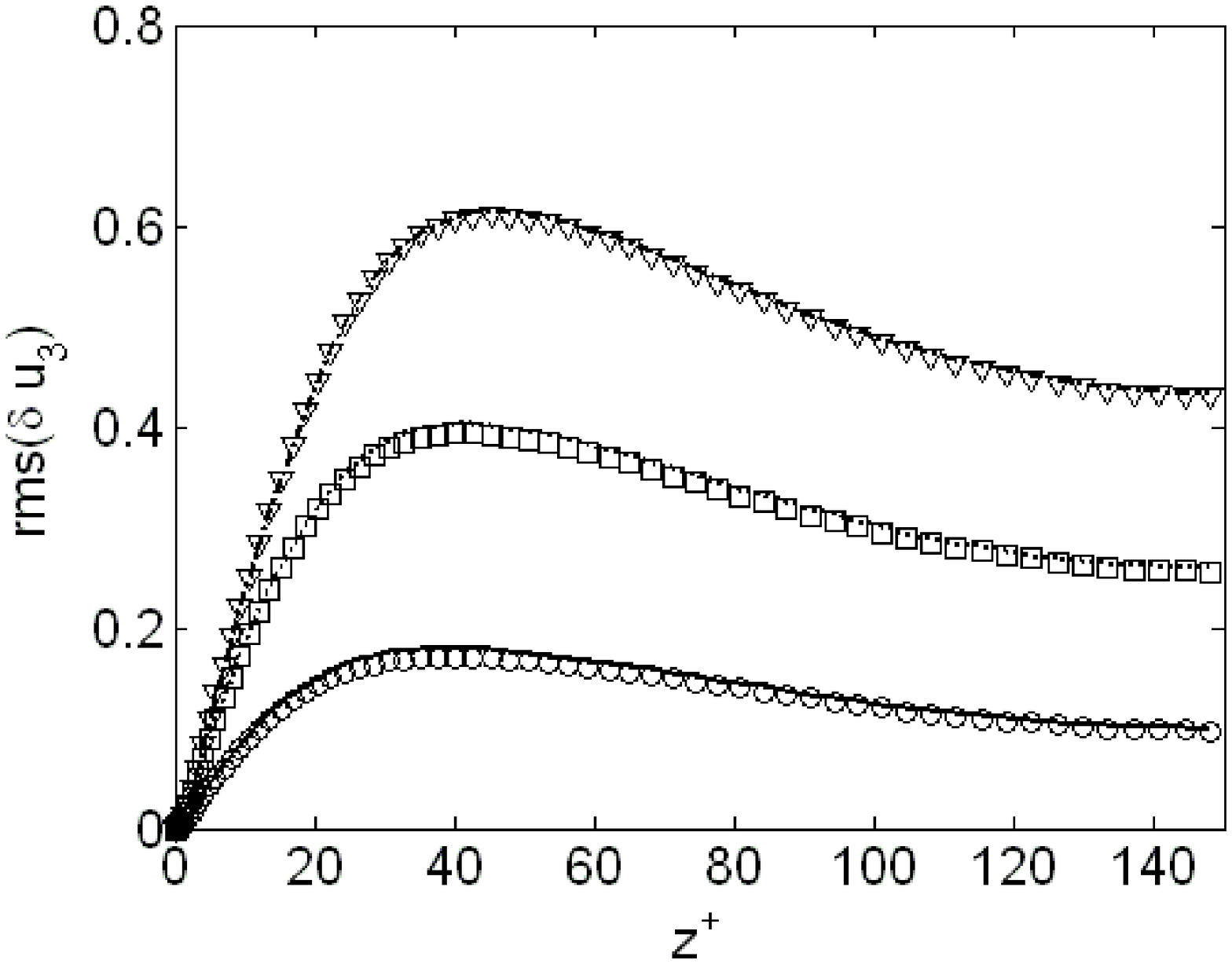}
\end{minipage}\hspace{2pc}
\begin{minipage}[t]{18pc}
\vspace{0pt}
\caption{\label{rms_IC_filter_z} R.m.s. values of the correction term component in the wall-normal direction as a function of $z^+$. St=5.}
\end{minipage}
\end{figure}

As for the r.m.s. of all the components of the correction term, reported in Figures \ref{rms_IC_filter_x}-\ref{rms_IC_filter_z}, their values increase with the filter width, as expected since for larger filter widths a larger amount of fluid velocity fluctuations is damped. There is no significant effect of the filter type.

\begin{figure}
\begin{minipage}{18pc}
\includegraphics[width=18pc,angle=0.]{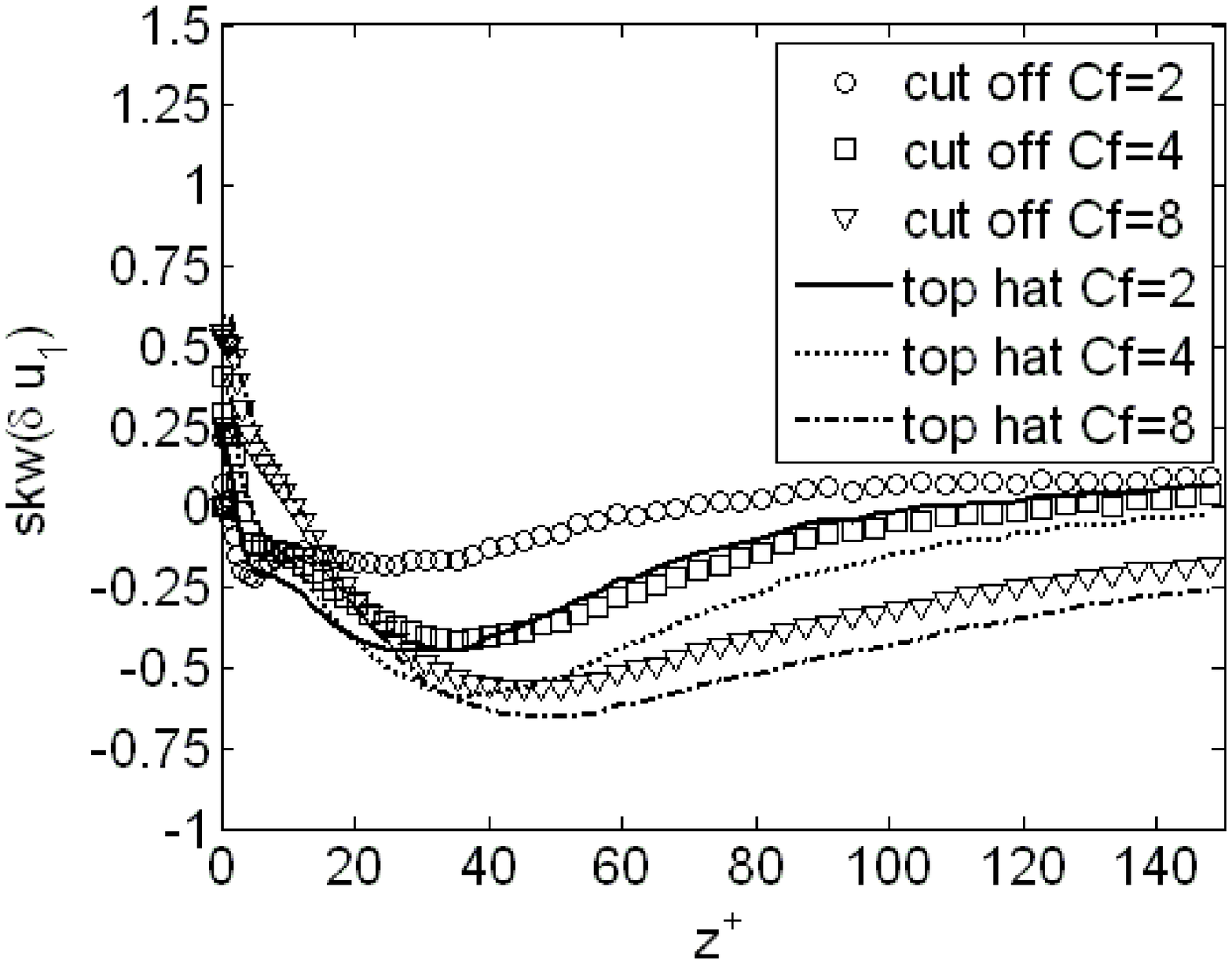}
\caption{\label{skw_IC_filter_x} Skewness of the correction term component in the streamwise direction as a function of $z^+$. St=5.}
\end{minipage}\hspace{2pc}
\begin{minipage}{18pc}
\includegraphics[width=18pc,angle=0.]{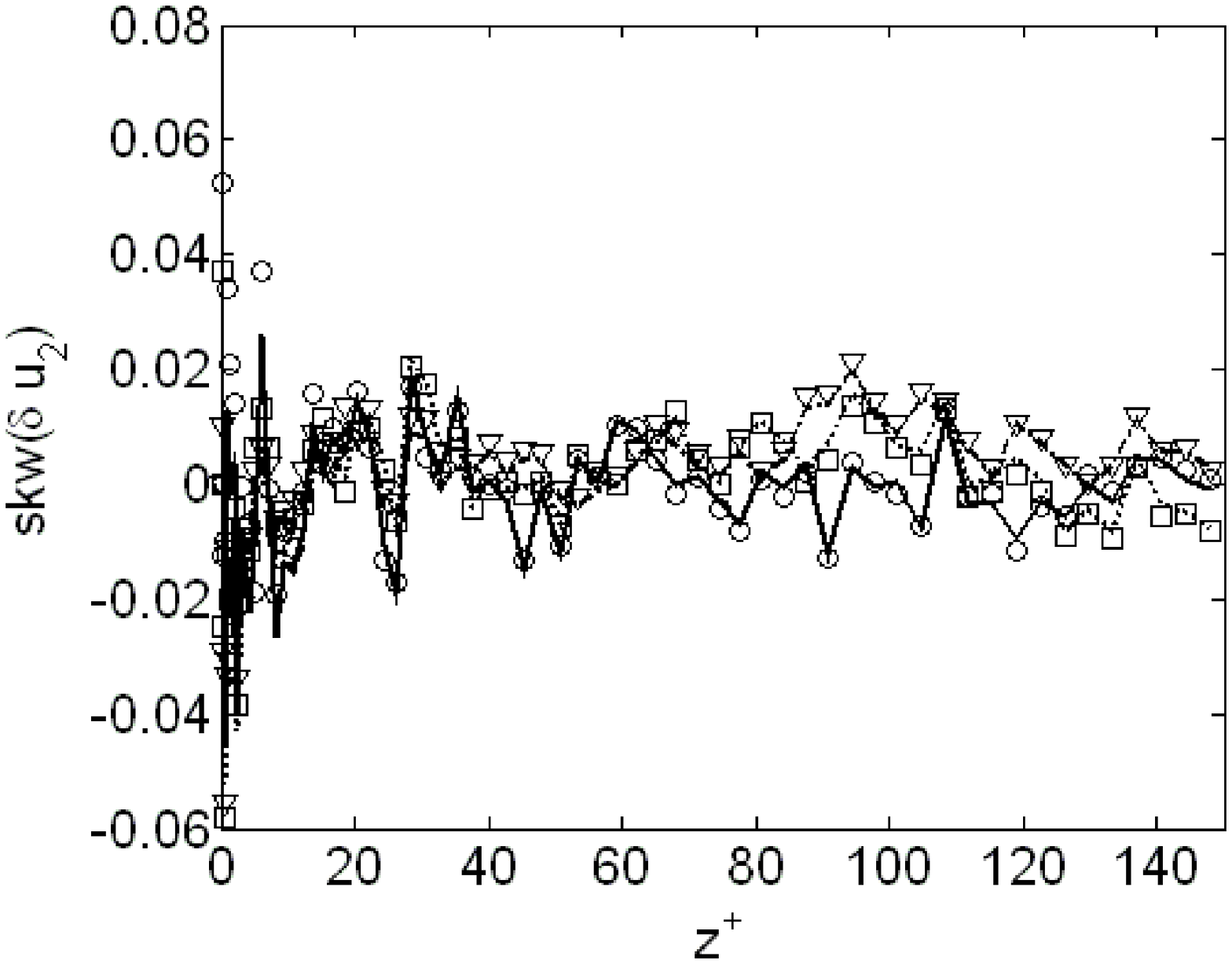}
\caption{\label{skw_IC_filter_y} Skewness of the correction term component in the spanwise direction as a function of $z^+$. St=5.}
\end{minipage}\hspace{2pc}
\begin{minipage}[t]{18pc}
\vspace{0pt}
\includegraphics[width=18pc,angle=0.]{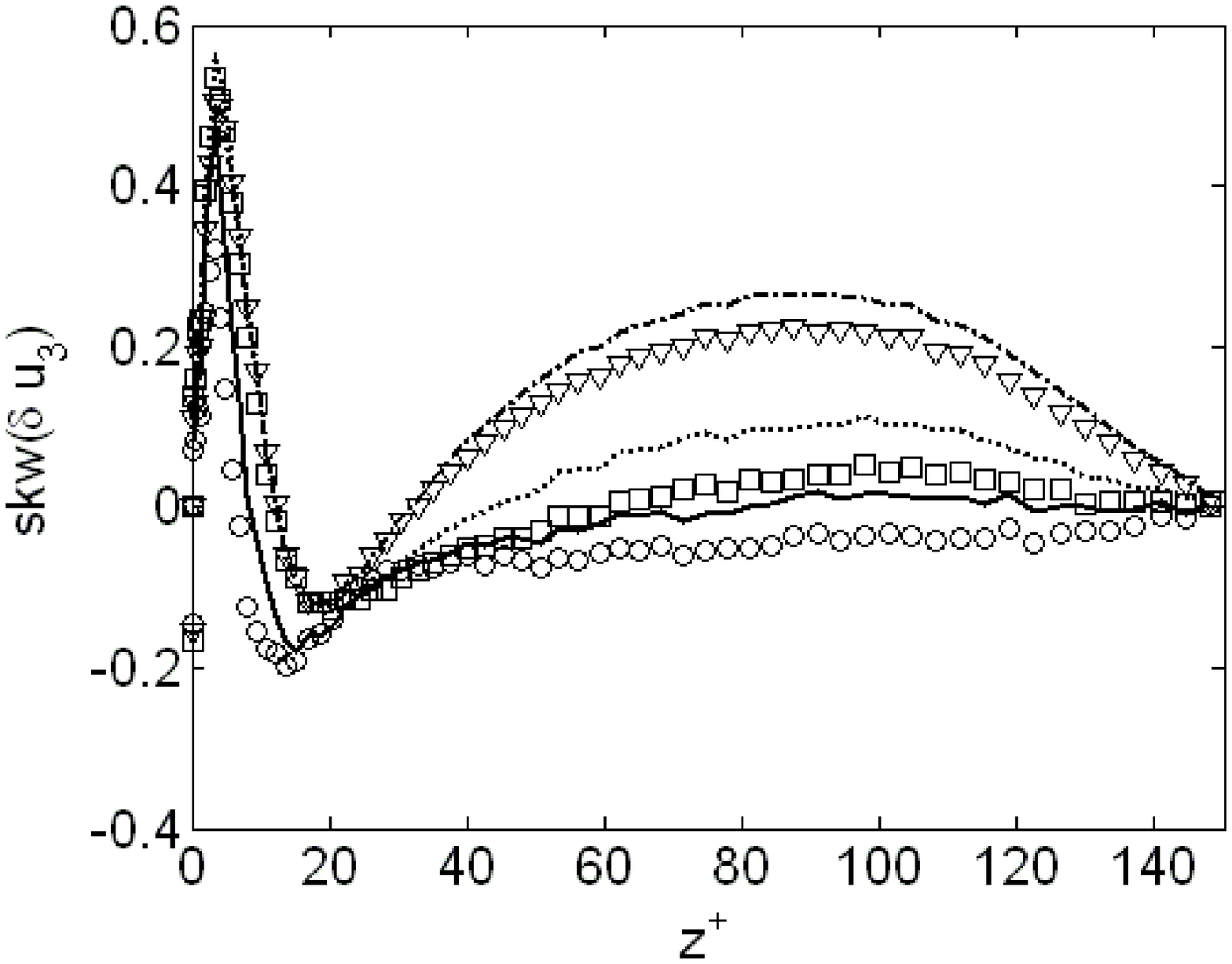}
\end{minipage}\hspace{2pc}
\begin{minipage}[t]{18pc}
\vspace{0pt}
\caption{\label{skw_IC_filter_z} Skewness of the correction term component in the wall-normal direction as a function of $z^+$. St=5.}
\end{minipage}
\end{figure}

The skewness of the different components of the SGS correction term is shown in Figs. \ref{skw_IC_filter_x}-\ref{skw_IC_filter_z}. As observed in Sec. \ref{statistics_inertia}, the values for the spanwise component are practically negligible for all the considered filters. Significant effects of the filter width, and in a less extent of the filter type, are conversely observed for both the streamwise and wall-normal components, although the qualitative behaviors remain the same. 

\begin{figure}
\begin{minipage}{18pc}
\includegraphics[width=18pc,angle=0.]{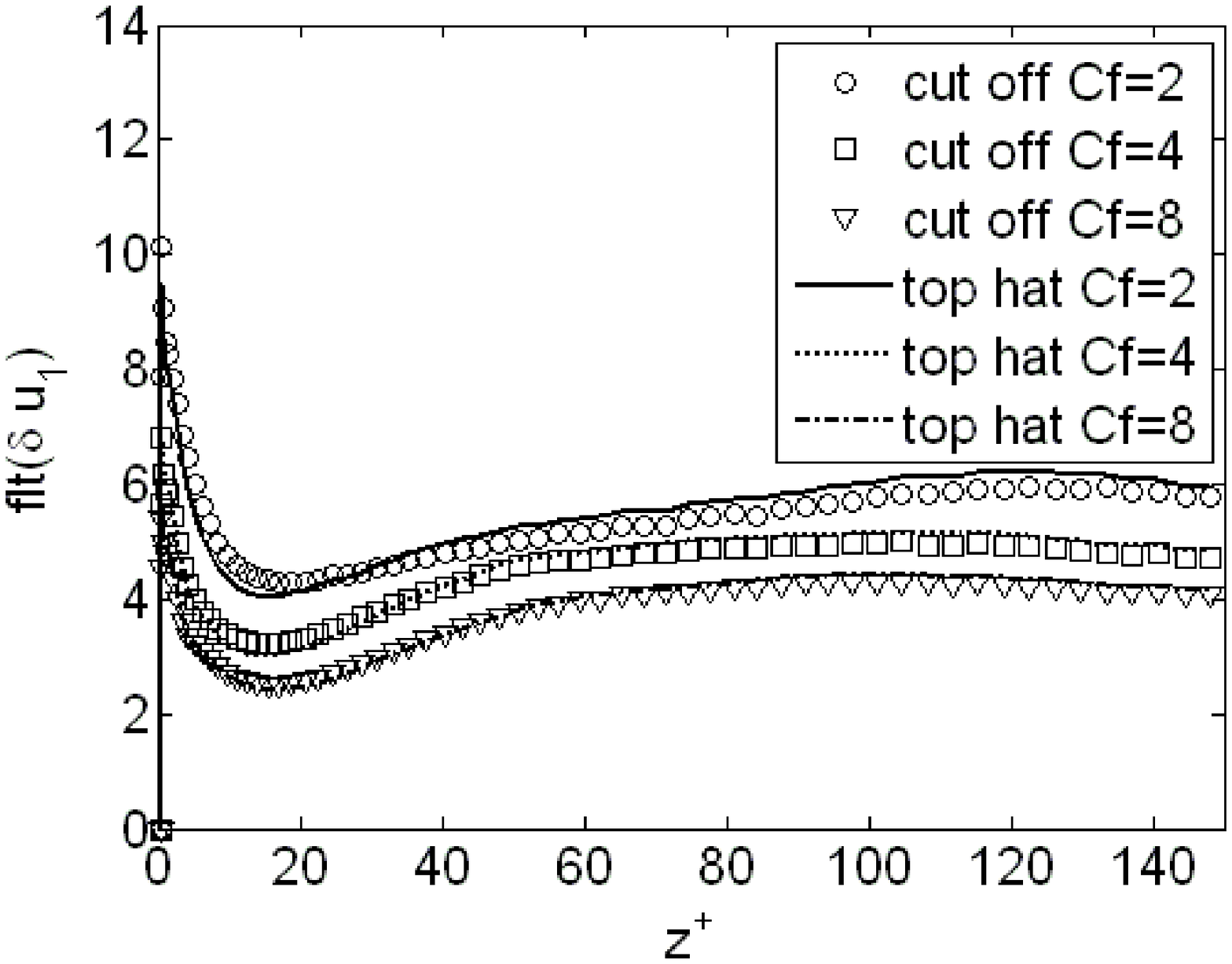}
\caption{\label{flt_IC_filter_x} Flatness of the correction term component in the streamwise direction as a function of $z^+$. St=5.}
\end{minipage}\hspace{2pc}
\begin{minipage}{18pc}
\includegraphics[width=18pc,angle=0.]{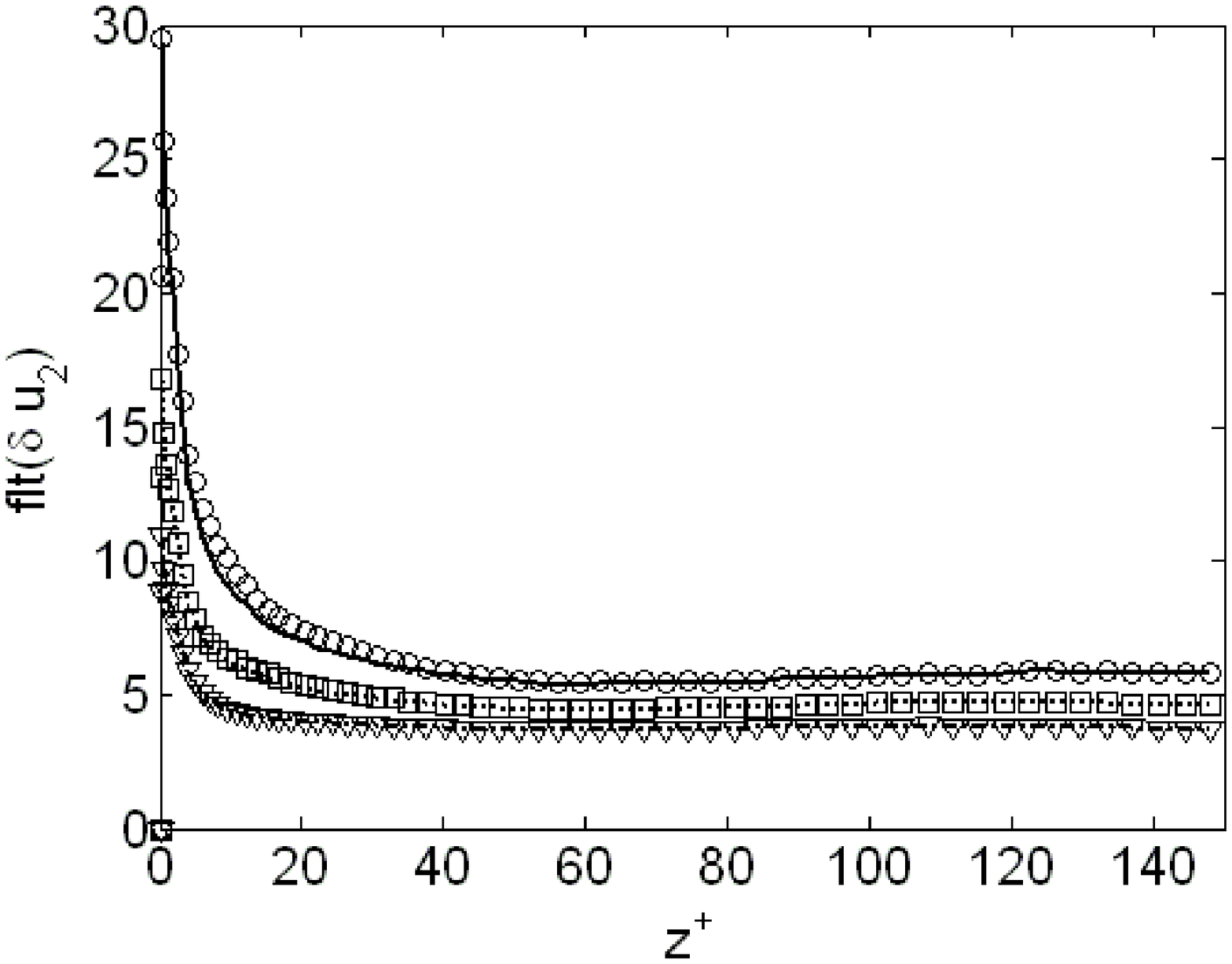}
\caption{\label{flt_IC_filter_y} Flatness of the correction term component in the spanwise direction as a function of $z^+$. St=5.}
\end{minipage}\hspace{2pc}
\begin{minipage}[t]{18pc}
\vspace{0pt}
\includegraphics[width=18pc,angle=0.]{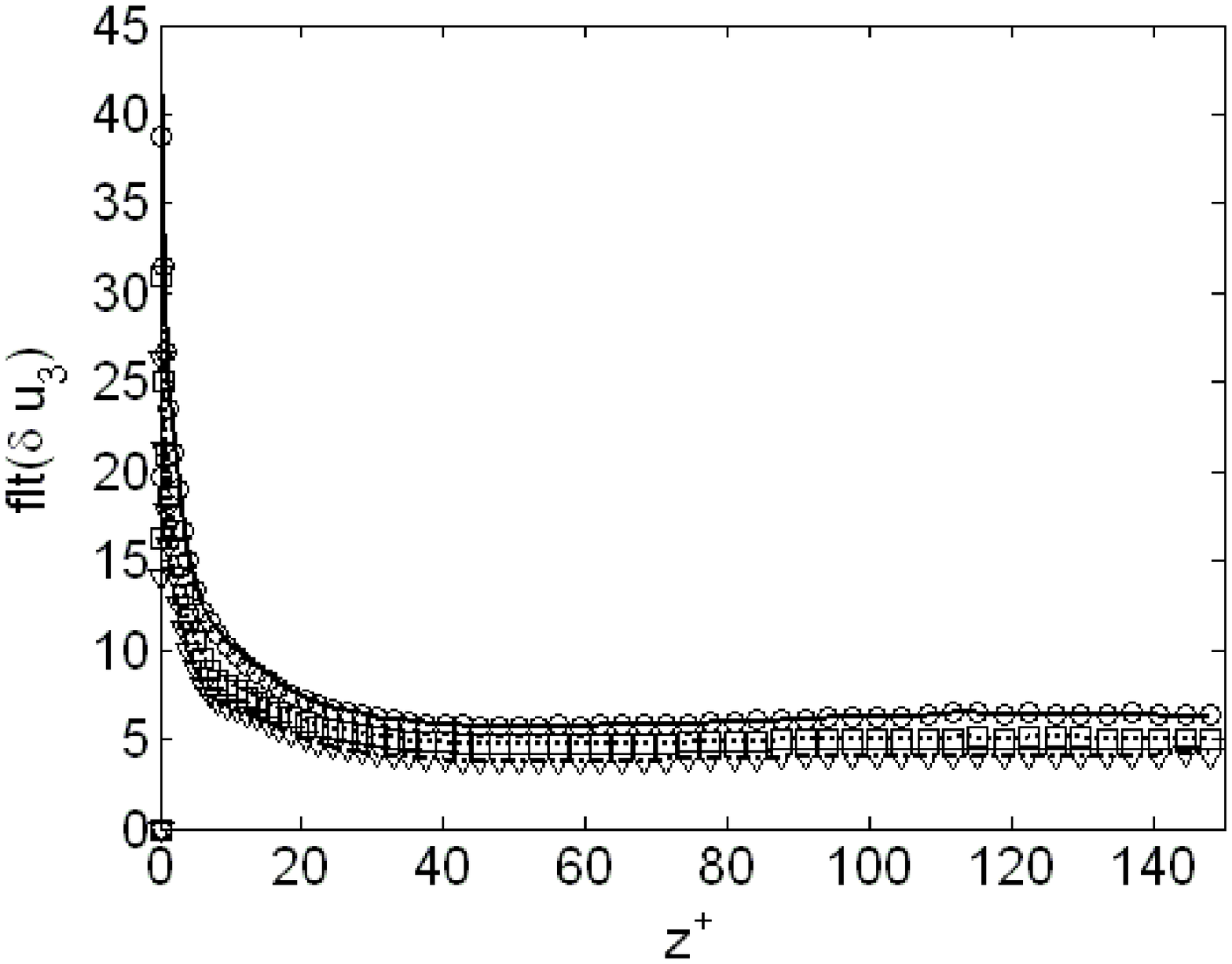}
\end{minipage}\hspace{2pc}
\begin{minipage}[t]{18pc}
\vspace{0pt}
\caption{\label{flt_IC_filter_z} Flatness of the correction term component in the wall-normal direction as a function of $z^+$. St=5.}
\end{minipage}
\end{figure}

Finally, the SGS velocity correction flatness slightly tends to decrease when increasing the filter width (Figs. \ref{flt_IC_filter_x}-\ref{flt_IC_filter_z}), but the behavior remains qualitatively the same for all the considered cases as described in Sec. \ref{statistics_inertia}.

\subsection{Concluding Remarks}

The pure effect of filtering on the fluid velocity seen by inertial heavy particles has been singled out by imposing that the particle trajectories in a-priori LES of turbulent channel flow coincide with those computed in DNS. An ideal SGS correction term has therefore been computed for different particle inertia and different filter types and widths. The statistical moments of this correction term have been investigated.

The first surprising result is that the mean values of the correction term components in the streamwise and wall-normal directions are not zero, as it would be expected from an Eulerian analysis, i.e. by looking at filtering effects on the mean value of fluid velocity at fixed points. Indeed, the ideal correction takes into account the effects of filtering on the fluid velocity seen by the particles, i.e. computed along the particle trajectories and the non-zero mean values depend on the effects of filtering on the near-wall turbulence structures combined with particle preferential sampling.  As for the streamwise component, the correction term in mean tends to reintroduce the effects of low-speed streaks preferentially sampled by the particles and smoothed by filtering; therefore, it is characterized by a negative peak in the near wall region. The quantitative value of this near-wall peak depends on both particle inertia and filter width. As for the mean wall-normal component, the behavior of the ideal correction term is more complex, but again tends to compensate filtering effects on the near-wall structures, which in turn affect ejection events, preferentially sampled by inertial particles. Particle inertia and filter width also affect the qualitative behavior of this component of the mean SGS correction. In spite of this different and complex behavior, as a general remark, the error introduced by filtering on inertial particle motion is not only due to the reduction of fluid velocity fluctuations, which is a well known effect pointed out in classical error analysis in LES carried out for fixed points. This supports the conclusion drawn in our previous \cite{Marchioli2008,Marchioli2008b}, i.e. that the reintroduction of the correct amount of fluid velocity fluctuations by a SGS model seems not to be enough to have an accurate prediction of particle preferential concentration and near-wall accumulation. 

By analyzing the higher-order moments of the correction term, the pdf of the SGS correction term is expected to significantly deviate from a Gaussian distribution. This deviation seems to be more important near the wall and the shape of the pdf is expected to be different for the different components of the ideal correction. Conversely, the effects on the pdf shape of the filter width and type, and even more of particle inertia, are expected to be small. As previousl mentioned, the Eulerian and Lagrangian pdfs of the ideal correction term will be presented and analyzed in forthcoming studies.

It would be also interesting to compute the ideal SGS correction for different values of the Reynolds number, in order to investigate whether Reynolds number effects should be taken into account.
\section*{Acknowledgments}
CINECA supercomputing center (Bologna, Italy) is gratefully acknowledged for generous allowance of computer resources. The authors wish to thank Prof. B.J. Geurts for the interesting discussion which stimulated the present study.
\section*{References}
\bibliographystyle{iopart-num}
\bibliography{Salvetti_etal}
\end{document}